\documentclass[a4paper, 11pt]{article}
\usepackage[utf8]{inputenc}
\usepackage{jcappub}
\usepackage{amsfonts}
\usepackage{graphicx}
\usepackage{amssymb}
\usepackage{amsmath}
\usepackage{breqn}
\usepackage{array}

\title{Complete Hamiltonian analysis of cosmological perturbations at
  all orders}

\author{Debottam Nandi} \author{and S. Shankaranarayanan}
\affiliation{School of Physics, Indian Institute of Science Education
  and Research Thiruvananthapuram (IISER-TVM), India}

\emailAdd{debottam@iisertvm.ac.in}
\emailAdd{shanki@iisertvm.ac.in}

\abstract{In this work, we present a consistent Hamiltonian analysis
  of cosmological perturbations at all orders. To make the procedure
  transparent, we consider a simple model and resolve the
  `gauge-fixing' issues and extend the analysis to scalar field models
  and show that our approach can be applied to any order of
  perturbation for any first order derivative fields. In the case of
  Galilean scalar fields, our procedure can extract constrained
  relations at all orders in perturbations leading to the fact that
  there is no extra degrees of freedom due to the presence of higher
  time derivatives of the field in the Lagrangian. We compare and
  contrast our approach to the Lagrangian approach (Chen et al [2006]) 
  for extracting higher order correlations and show
  that our approach is efficient and robust and can be applied to any
  model of gravity and matter fields without invoking slow-roll approximation.}
\begin{document}
\maketitle

\section{Introduction}
Linear order cosmological perturbation
theory~\cite{Bardeen:1980kt,Langlois1994,Mukhanov:1990me,Sasaki1986,Lidsey:1995np,Garriga1999,Kodama01011984}
has been highly successful at describing the CMB
anisotropies~\cite{Komatsu:2008hk}. It also helps to describe the seed
Gaussian density perturbations during inflation and the formation of
large scale
structures~\cite{STAROBINSKY1982175,BardeenSteinhardt1983}. While
linear order perturbations are fairly understood, there are several
open issues in applying theory beyond linear order both early and late
time
universe~\cite{Salopek1990,STAROBINSKY1982175,Sasaki:1995aw,Lyth:1998xn,Lidsey:1995np,Armendariz-Picon1999,Garriga1999}. In
the last decade, the possibility of observing primordial
non-Gaussianity in CMB\cite{Komatsu:2008hk} and potentially ruling out
inflationary
models\cite{Lidsey:1995np,Lyth:1998xn,Lyth:2007qh,Mazumdar:2010sa}
has led to a lot of interest in higher order perturbations.

Currently, there are two formalism in the literature to study gauge
invariant cosmological perturbations --- 
Hamiltonian~\cite{Langlois1994} and Lagrangian
formulation~\cite{Bardeen:1980kt,Mukhanov:1990me,Bruni:1996im,Acquaviva:2002ud,Nakamura:2003wk,PhysRevD.69.104011,Bartolo:2001cw,Rigopoulos:2002mc,Bernardeau:2002jy,Bernardeau:2002jf,Malik:2003mv,
  Bartolo:2003bz, Finelli:2003bp,Bartolo:2004if,Enqvist:2004bk,Vernizzi:2004nc,Tomita:2005et,
  Lyth:2005du, Seery:2005gb, Malik:2005cy,Seery:2006vu,Lifshitz:1963ps,Lukash:1980iv}. In the Lagrangian
formulation, one needs to either perturb Einstein's equations or vary
the perturbed Lagrangian to obtain perturbed equations of motion. In
the Hamiltonian formulation, gauge-invariant first order perturbed
equations are obtained in terms of field variables and their conjugate
momenta.
  
In the context of early universe, evaluation of $n$-point correlation
functions of the effective (scalar) field requires the quantum
Hamiltonian of this effective field. For matter fields containing
first derivative in time, it is straightforward to obtain the
Hamiltonian by performing Legendre transformation. Equations of motion
of such fields contain upto second derivative of the field variables
which can be linearized as two independent coupled first order
differential equations (Hamilton's equations) --- one corresponding to
the time evolution of the field and other corresponding to the time
evolution of the momentum. This indicates that the phase space is
two-dimensional.

However, for higher derivative field theories, the Hamiltonian
structure and the associated degrees of freedom is not
straightforward. For any higher (more than one) derivative theories,
the equations of motion have upto two times the highest order
derivative of the field. For example, fields with second order time
derivatives, the equation motion contain upto fourth order derivatives
of the field. So, if we linearize the equation, we obtain four
independent coupled first order differential equations which
indirectly imply that, the phase-space is four dimensional and can be
mapped to Hamiltonian of two fields. However, the mapped Hamiltonian
has unbounded negative energy leading to Ostrogradsky's
instability~\cite{Ostro,Woodard:2015zca}. This implies that extra
degrees of freedom (named as ghost) for a higher derivative Lagrangian
causes the instability and hence, in general, quantizing the
Hamiltonian is not possible.

On the contrary, Galilean scalar
field~\cite{Kobayashi2010,Kobayashi2011,Deffayet2009} is a special
higher derivative field which leads to second derivative equations of
motion, implying that the the phase space contains one independent
variable and one corresponding momentum, although, multiple variable as
well as momenta may appear in the Hamiltonian. Also the absence of
extra degrees of freedom leads to the fact that, Hamiltonian of
Galilean field is bounded can be quantized.

The main aim of the work is to write the effective Hamiltonian of the
generalized (Galilean) scalar field coupled to gravity at all orders
in perturbations. Hamiltonian approach to the cosmological
perturbations has not been extensively studied in the
literature. Langlois~\cite{Langlois1994} showed that, equations of
motion of canonical scalar field can be obtained in a gauge-invariant
single variable form at first order perturbation. However, Langlois'
approach can not be extended to include higher order due to fact that
the approach requires construction of gauge-invariant conjugate
momentum. Another aim of this work is to extend Langlois' analysis to
higher orders. It is necessary to extend Langlois' method to higher
order for the following reasons. First, to calculate higher order
correlation functions, currently several approximations are employed
to convert effective Lagrangian to
Hamiltonian~\cite{Huang:2006eha}. Our aim is to provide a simple, yet
robust procedure to calculate Hamiltonian for arbitrary field(s) at
all orders in cosmological perturbations. Second, as mentioned
earlier, we do not have a procedure to perform Hamiltonian analysis
for the Galilean fields. The procedure we adopt here can be extended
to Galilean fields; we explicitly show this in this work. Deffayet et
al~\cite{Deffayet:2015qwa} gave a mechanism to deal with Galilean
theory in the context of General relativity. Third, the procedure can
be used to include quantum gravitational
corrections\cite{Barrau:2013ula,Ashtekar:2013xka,Bojowald:2012xy}.

In this work, we find a consistent perturbed Hamiltonian
formulation. We use Deffayet's approach~\cite{Deffayet:2015qwa} to
obtain the generalized Hamiltonian of a Galilean theory and along with
canonical scalar field Hamiltonian, we perturb both fields to obtain
all equations of motion as well as interaction Hamiltonian and we
compare with conventional Lagrangian formulation. We find that both
lead to identical results and hence, our Hamiltonian approach leads to
consistent results in a straightforward and efficient way.

In section \ref{gauge}, we introduce the generic scalar model in the early Universe and briefly discuss gauge fixing and the corresponding gauge invariant equations of motion. In section \ref{simple model}, we 
take a simple model that highlights the key issues that need to be addressed about the gauge issue in Hamiltonian formulation and also discuss how the same can be addressed. In this simple model, we compare and contrast Lagrangian and Hamiltonian formulation. We calculate third and fourth order perturbed Hamiltonian and replace momenta with the time derivatives of the variables and show that it is consistent with conventional Lagrangian formulation. In section \ref{EoMLC}, we discuss canonical scalar field in flat-slicing gauge to obtain interaction Hamiltonian in a new and
simple way. In Appendix \ref{PLeCSF}, we calculate equations of motion of perturbed and unperturbed variables for Canonical scalar field in flat-slicing gauge. We also explicitly obtain the third order interaction Hamiltonian of Canonical scalar field model in phase-space. In order to show that our proposed method works in any gauges, we obtain equations of motion of all variables of Canonical scalar field in uniform density gauge in Appendix \ref{udgcsf}. In section
\ref{galileanfield} and Appendix \ref{galmod}, we extend the analysis to a very specific Galilean field and evaluate all equations of motion of
perturbed-unperturbed variables. We show that, at every order unlike
higher order generalized Lagrangian, Galilean field model does not
provide extra degrees freedom and behave same as any general first
order derivative Lagrangian model. We also calculate the third and
fourth order perturbed Hamiltonian. In Appendix \ref{galcan}, we
consider a Galilean model with a canonical scalar field part and
express the full Hamiltonian as well as zeroth and second order
perturbed Hamiltonian and express zeroth and first order perturbed
equations.

In this work, we consider $(-, +, + ,+)$ metric signature. We also
denote ${}^\prime$ as derivative with respect to conformal time.

\section{Basic models and Gauge choices}\label{gauge}
Action for a generic scalar field $(\varphi)$ minimally coupled to gravity is
\begin{equation}
\label{GravityAction}
\mathcal{S} = \int d^4 x \sqrt{-g} \left[\frac{1}{2 \kappa} R + \mathcal{L}_m(\varphi, \partial\varphi, \partial\partial\varphi) \right],
\end{equation}
where $R$ is the Ricci scalar and the matter Lagrangian, $\mathcal{L}_m$ is of the form.
\begin{equation}\label{boxx}
  \mathcal{L}_m = P(X, \varphi) + G(X, \varphi) \Box \varphi,~~~ ~X \equiv 
  \frac{1}{2} g^{\mu \nu} \partial_{\mu}{\varphi} \partial_{\mu}{\varphi}, ~~~ 
  \Box \equiv -\frac{1}{\sqrt{-g}}\partial_{\mu}\left(\sqrt{-g} g^{\mu \nu} \partial_{\nu}\right).
\end{equation}

It is important to note that the Lagrangian contains
second order derivatives, however, the equation of motion will be of
second order and these are referred as
Galilean\cite{Kobayashi2010,Kobayashi2011,Deffayet2009}. Varying the
action (\ref{GravityAction}) with respect to metric gives Einstein's
equation
\begin{equation}
\label{EinsteinEquation}
R_{\mu \nu} - \frac{1}{2} g_{\mu \nu}~ R = \kappa~ T_{\mu \nu}, 
\end{equation}
where the stress tensor $T_{\mu \nu}$ is
\begin{equation}
\label{EMTensor}
\begin{split}
T_{\mu \nu} &= g_{\mu \nu} \left( P + G_X g^{\alpha
  \beta} \partial_{\alpha}{X} \partial_{\beta}{\varphi} + G_\varphi
g^{\alpha
  \beta} \partial_{\alpha}{\varphi} \partial_{\beta}{\varphi}\right) \\
 & - \left(P_X + 2 G_\varphi + G_X \Box
\varphi\right) \partial_{\mu}{\varphi} \partial_{\nu}{\varphi} - 2
G_X \partial_{\mu}{X} \partial_{\nu}{\varphi}.
\end{split}
\end{equation}

Varying the action (\ref{GravityAction}) with respect to the scalar
field `$\varphi$' leads to the following equation of motion
\begin{equation}
\label{EG}
\begin{split}
  &\left(2 G_\varphi - 2 X G_{X \varphi} + P_X\right) \Box \varphi - \left(P_{XX} +
  2 G_{X \varphi}\right)\partial_{\mu}{\varphi} \partial^{\mu}{\varphi} -
  2X
  \left(G_{\varphi} + P_{X \varphi} \right) + P_\varphi\\
  & - G_X \left( \varphi_{, \mu \nu} \varphi_{,}^{\mu \nu} - \left(\Box
  \varphi\right)^2 + R_{\mu
    \nu}\partial^{\mu}{\varphi} \partial^{\nu}{\varphi}\right) - G_{XX}
 \left( \partial_{\mu}X \partial^{\mu}X +
  \partial_{\mu}{\varphi} \partial^{\mu}X\,\Box \varphi\right) = 0,
\end{split}
\end{equation}
which can also be obtained by using the conservation of
Energy-Momentum tensor, $\nabla_{\mu}{T^{\mu \nu}} = 0$. Setting $G(X,
\varphi) = 0$ corresponds to non-canonical scalar field. Further,
fixing $ P = - X - V(\varphi)$, where $V(\varphi)$ is the potential,
corresponds to canonical scalar field.

The four-dimensional line element in the ADM form is given by,
\begin{eqnarray}
ds^2 &=& g_{\mu \nu} dx^{\mu} dx^{\nu} \nonumber \\
\label{line}
&=& -(N^2 - N_{i} N^{i} ) d\eta^2 + 2 N_{i} dx^{i} d\eta + \gamma_{i j} dx^{i} dx^{j},
\end{eqnarray}
where $N(x^{\mu})$ and $N_i(x^{\mu})$ are Lapse function and Shift
vector respectively, $\gamma_{i j}$ is the 3-D space metric.  Action
(\ref{GravityAction}) for the line element (\ref{line}) takes the
form,
\begin{equation}
\label{AA}
\mathcal{S} = \int d^4x \sqrt{-g} \Big\{ \frac{1}{2 \kappa}
\left(^{(3)}R + K_{i j} K^{i j} - K^2\right) + \mathcal{L}_m \Big\}
\end{equation}
where $K_{i j}$ is extrinsic curvature tensor and is defined by
\begin{eqnarray}
  && K_{i j} \equiv \frac{1}{2N} \Big[ \partial_{0}{\gamma_{i j}} - N_{i|j} - N_{j | i} \Big] \nonumber \\
  && K \equiv \gamma^{i j} K_{i j} \nonumber
\end{eqnarray}

Perturbatively expanding the metric only in terms of scalar perturbations and the scalar field  about the
flat FRW spacetime in conformal coordinate, we get,

\begin{eqnarray}
  && g_{0 0} = - a(\eta)^2(1 + 2 \epsilon \phi_1 + \epsilon^2 \phi_2 + ...) \\
  && g_{0 i} \equiv N_{i} = a(\eta)^2 (\epsilon \partial_{i}{B_1} + \frac{1}{2} \epsilon^2 \partial_{i}{B_2} + ...) \\
  && g_{i j} =a(\eta)^2 \left((1 - 2 \epsilon \psi_1 - \epsilon^2 \psi_2 -...)\delta^{i j} +  2 \epsilon E_{1 i j} + \epsilon^2 E_{2 i j} + ...\right)\\
  &&\varphi = \varphi_0(\eta) + \epsilon \varphi_1 + \frac{1}{2} \epsilon^2 \varphi_2 + ...
\end{eqnarray}
where $\epsilon$ denotes the order of the perturbation.
To determine the dynamics at every order, we need five scalar functions ($\phi, B,
\psi, E$ and $\varphi$) at each order. Since there are two free gauge
choices, one can fix two of the five scalar functions. In this work,
we derive all equations by choosing a specific gauge --- flat-slicing
gauge, i.e., $\psi = 0, E = 0$ --- at all orders:
\begin{eqnarray}
\label{00pmetric}
  && g_{0 0} =- a(\eta)^2(1 + 2 \epsilon \phi_1 + \epsilon^2 \phi_2 + ...) \\
  \label{0ipmetric}
  && g_{0 i} \equiv N_{i} = a(\eta)^2 (\epsilon \partial_{i}{B_1} + \frac{1}{2} \epsilon^2 \partial_{i}{B_2} + ...) \\
  \label{3metric}
  && g_{i j} =a(\eta)^2 \delta_{i j}\\
  \label{field}
  &&\varphi = \varphi_0(\eta) + \epsilon \varphi_1 + \frac{1}{2} \epsilon^2 \varphi_2 + ...
\end{eqnarray}

It can be shown that, perturbed equations in flat-slicing
  gauge coincide with gauge-invariant equations of motion (in generic
  gauge, $\varphi_1$ coincides with $\varphi_1 +
  \frac{\varphi_0{}^\prime}{H} {}\psi_1 \equiv
  \frac{\varphi_0{}^\prime}{H} \mathcal{R}$ which is a
  gauge-invariant quantity, $\mathcal{R}$ is called curvature
  perturbation). Similarly, one can choose another suitable gauge with
  no coordinate artifacts to obtain gauge-invariant equations of
  motion\cite{Malik2009}. Such gauges are Newtonian-conformal gauge
  $(B = 0,\, E = 0)$, constant density gauge $(E = 0, \, \delta\varphi
  = 0)$, etc. 
  
  One immediate question that needs to be addressed in the Hamiltonian 
  formulation is the following: for a given gauge choice, if a particular 
  set of variables are set to zero, whether the corresponding conjugate 
  momenta also vanish? In other words, in the flat-slicing gauge $\delta g_{ij} = 0$, 
  does this mean the corresponding canonical conjugate momentum $\delta \pi^{i
    j}$ vanish? In order to go about understanding this, in the next section, we take a simple 
    model of two variables ($x$ and $y$) where one of the variables is perturbed, 
    while the other variable is not perturbed and study the Hamiltonian formulation of this model.

\section{Simple model: Warm up}\label{simple model}

As discussed above, in cosmological perturbation theory, by fixing a `gauge', 
we assume some field variables to be unperturbed where some variables are
perturbed. To go about understanding the procedure in the Hamiltonian
formulation, we consider a simple classical model that consists of
both perturbed and unperturbed variables. We also show that the
Hamilton's equations of unperturbed as well as perturbed variables are
identical to Euler-Lagrange equations of motion. The Lagrangian of the
simple model is
\begin{equation}
\label{fulllagrangian}
\mathcal{L} = \frac{1}{2 y}\, \left((\partial_t x){}^2 + (\partial_ty){}^2\right) - \frac{1}{4}\left(x{}^{4} +  y{}^{4}\right).
\end{equation}
The corresponding momenta are
\begin{equation}
\label{momenta}
\pi_x\, = \,\frac{\partial_tx}{y} ,~~\pi_y\,=\,\frac{\partial_ty}{y},
\end{equation}
and the corresponding Hamiltonian (\ref{fulllagrangian}) is given by
\begin{equation}
\label{fullhamiltonian}
\mathcal{H} = \frac{1}{2} y \left(\pi_x{}^2 + \pi_y{}^2\right) + \frac{1}{4} \left(x^4 + y^4\right).
\end{equation}

\subsection{Perturbed Lagrangian}
As mentioned earlier, we consider $x$ to be unperturbed and 
$y$ to be perturbed, and separate into background and perturbed parts, i.e.,
\begin{equation}
\label{xy}
x = x_0, ~~~y = y_0 + \epsilon\, y_1.
\end{equation}
where $\epsilon$ is the order of perturbation. In this work, we mainly
focus on first order perturbation, however, the analysis can be
extended to any higher order perturbations. Using (\ref{xy}), we
separate the Lagrangian (\ref{fulllagrangian}) into a background part
and perturbed parts, and write it as
\begin{equation}
\mathcal{L} = \mathcal{L}_0 + \epsilon \mathcal{L}_1 + \epsilon^2 \mathcal{L}_2 + \epsilon^3 \mathcal{L}_3 + \epsilon^4 \mathcal{L}_4 + ...
\end{equation}
where
\begin{eqnarray}
\label{PerturbedLagrangian}
\label{zerothL}
&&\mathcal{L}_0 = \frac{1}{2}\, (\partial_t{x_0}\, ){}^{2} y_0{}^{-1} + \frac{1}{2}\, (\partial_t{y_0}\, ){}^{2} y_0{}^{-1} - \frac{1}{4}\, \left(x_0{}^{4} + y_0{}^{4}\right)  \\
&& \mathcal{L}_1 = \partial_t{x_0}\,  \partial_t{y_1}\,  y_0{}^{-1} - \frac{1}{2}\, (\partial_t{x_0}\, ){}^{2} y_0{}^{(-2)} y_1 - \frac{1}{2}\, (\partial_t{y_0}\, ){}^{2} y_0{}^{(-2)} y_1 - y	_0{}^{3} y_1  \\
\label{secondL}
&&\mathcal{L}_2 = \frac{1}{2}\, (\partial_t{y_1}\, ){}^{2} y_0{}^{-1} - \partial_t{y_0}\,  \partial_t{y_1}\,  y_0{}^{(-2)} y_1 + \frac{1}{2}\, (\partial_t{x_0}\, ){}^{2} y_0{}^{(-3)} y_1{}^{2} + \nonumber \\
&&~~~~~~\frac{1}{2}\, (\partial_t{y_0}\, ){}^{2} y_0{}^{(-3)} y_1{}^{2} - \frac{3}{2}\, y_0{}^{2} y_1{}^{2}  \\
\label{thirdL}
&&\mathcal{L}_3 = - \frac{1}{2}\, (\partial_t{y_1}\, ){}^{2} y_0{}^{(-2)} y_1 + \partial_t{y_0}\,  \partial_t{y_1}\,  y_0{}^{(-3)} y_1{}^{2} - \frac{1}{2}\, (\partial_t{x_0}\, ){}^{2} y_0{}^{(-4)} y_1{}^{3} -\nonumber \\
&&~~~~~~\frac{1}{2}\, (\partial_t{y_0}\, ){}^{2} y_0{}^{(-4)} y_1{}^{3} - y_1{}^{3} y_0  \\
\label{fourthL}
&&\mathcal{L}_4 = \frac{1}{2}\, (\partial_t{y_1}\, ){}^{2} y_0{}^{(-3)} y_1{}^{2} - \partial_t{y_0}\,  \partial_t{y_1}\,  y_0{}^{(-4)} y_1{}^{3} + \frac{1}{2}\, (\partial_t{x_0}\, ){}^{2} y_0{}^{(-5)} y_1{}^{4} + \nonumber \\
&&~~~~~~\frac{1}{2}\, (\partial_t{y_0}\, ){}^{2} y_0{}^{(-5)} y_1{}^{4} - \frac{1}{4}\, y_1{}^{4}.
\end{eqnarray}
Euler-Lagrange equations of motion of $x$ and $y$ for the Lagrangian (\ref{fulllagrangian}) are
\begin{eqnarray}
&& \partial_t\left(\frac{\partial_tx}{y}\right) + x^3 = 0 \\
\label{eqy}
&& \partial_t\left(\frac{\partial_t{y}}{y}\right) = - \frac{1}{2y^2} \left((\partial_t{x}){}^2 + (\partial_t{y}){}^2\right) - y^3
\end{eqnarray}
We perform order-by-order perturbation of the above equation using
(\ref{xy}) which can also be obtained by varying the perturbed
Lagrangian. Zeroth order equations of $x$, i.e., $x_0$ and $y$, i.e.,
$x_0$ are given by

\begin{eqnarray}
\label{zerothx}
&& \partial_t\left(\frac{\partial_tx_0}{y_0}\right) + x_0^3 = 0 \\
\label{zerothy}
&& \partial_t\left(\frac{\partial_t{y_0}}{y_0}\right) = - \frac{1}{2y_0{}^2} \left((\partial_t{x_0}){}^2 + (\partial_t{y_0}){}^2\right) - y_0^3.
\end{eqnarray}
One can either perturb the equation (\ref{eqy}) or vary the second
order Lagrangian (\ref{secondL}) with respect to $y_1$ to obtain first
order perturbed equation of motion of $y$ and is given by
\begin{equation}
\label{firstoy1}
\partial_t\left(\frac{\partial_ty_1}{y_0} - \frac{\partial_ty_0}{y_0{}^2} y_1\right) = \frac{1}{y_0{}^3}\left((\partial_t{x_0}){}^2 + (\partial_t{y_0}){}^2\right) y_1 - \frac{\partial_t y_0}{y_0{}^2}\partial_ty_1 - 3y_0{}^2 y_1.
\end{equation}
In the next subsection, we explicitly write down the perturbed
equations using Hamiltonian (\ref{fullhamiltonian}).

\subsection{Perturbed Hamiltonian}
It is important to note that even though $x$ is not perturbed, $\pi_x$
contains both perturbed and unperturbed parts.  Using (\ref{xy}), the
following relations can easily be established\footnote{It is apparent from
  equations (\ref{momenta}) that only first order perturbation of $y$
  can produce any higher order perturbed momenta of $x$ and $y$ and it
  is given by
\begin{eqnarray}
\label{pmomx}
&&\pi_x = \pi_{x0} + \epsilon \pi_{x1} + \epsilon^2 \pi_{x2} ... \nonumber \\
\label{pmomy}
&&\pi_y = \pi_{y0} + \epsilon \pi_{y1} +  \epsilon^2 \pi_{y2} ...\nonumber
\end{eqnarray}
where
\begin{eqnarray}
\label{pmomentumx}
&&\pi_{x0} = \frac{d_tx_0}{y_0} ,~~ \pi_{x1} = - \frac{y_1}{y_0} \pi_{x0}, ~~ \pi_{x2} = - \frac{y_1}{y_0} \pi_{x1}, ~~ \pi_{x3} = -\frac{y_1}{y_0} \pi_{x2} ~~~...\nonumber \\
\label{pmomentumy}
&&\pi_{y0} = \frac{d_ty_0}{y_0}, ~~ \pi_{y1} = \frac{d_ty_1}{y_0} - \frac{y_1}{y_0} \pi_{y0}, ~~ \pi_{y2} = - \frac{y_1}{y_0} \pi_{y1}, ~~ \pi_{y3} = - \frac{y_1}{y_0} \pi_{y2} ~~ ...\nonumber
\end{eqnarray}
Since we are not interested in higher order perturbation theory, we neglect higher order momenta and consider only first order momenta in calculating correlation functions.}
\begin{eqnarray}
\label{pmomx}
&&\pi_x = \pi_{x0} + \epsilon {}\pi_{x1}\\
\label{pmomy}
&&\pi_y = {}\pi_{y0} + \epsilon {}\pi_{y1}
\end{eqnarray}
where
\begin{eqnarray}
\label{pmomentumx}
&&{}\pi_{x0} = \frac{\partial_tx_0}{y_0} ,~~ {}\pi_{x1} = - \frac{y_1}{y_0} {}\pi_{x0}\\
\label{pmomentumy}
&&{}\pi_{y0} = \frac{\partial_ty_0}{y_0}, ~~ {}\pi_{y1} = \frac{\partial_ty_1}{y_0} - \frac{y_1}{y_0} {}\pi_{y0}
\end{eqnarray}

Using (\ref{xy}), (\ref{pmomx}) and (\ref{pmomy}), Hamiltonian of the
system (\ref{fullhamiltonian}) can be written as
\begin{equation}
  \mathcal{H} = \mathcal{H}_0 + \epsilon \mathcal{H}_1 + \epsilon^2 \mathcal{H}_2 + \epsilon^3 \mathcal{H}_3 + \epsilon^4 \mathcal{H}_4 + ...
\end{equation}
where
\begin{eqnarray}
\label{PH0}
&&\mathcal{H}_0 = \frac{1}{2}\, y_0 \left({}\pi_{x0}{}^{2} +  {}\pi_{y0}{}^{2}\right) + \frac{1}{4}\, \left( x_0{}^{4} + y_0{}^{4}\right) \\
\label{PH1}
&&\mathcal{H}_1 = \frac{1}{2}\, {}\pi_{x0}{}^{2} y_1 + {}\pi_{x0} {}\pi_{x1} y_0 + \frac{1}{2}\, {}\pi_{y0}{}^{2} y_1 + {}\pi_{y0} {}\pi_{y1} y_0 + y_0{}^{3} y_1 \\
\label{PH2}
&&\mathcal{H}_2 = y_1 \left( {}\pi_{x0} {}\pi_{x1} + {}\pi_{y0} {}\pi_{y1}\right)  + \frac{1}{2}\, y_0 \left({}\pi_{x1}{}^{2} + {}\pi_{y1}{}^2\right)   + \frac{3}{2}\, y_0{}^{2} y_1{}^{2} \\
\label{PH3}
&& \mathcal{H}_3 =  \frac{1}{2}\, y_1\left({}\pi_{x1}{}^{2}  + {}\pi_{y1}{}^{2} \right)  + y_1{}^{3} y_0 \\
\label{PH4}
&& \mathcal{H}_4 = \frac{1}{4} y_1{}^4
\end{eqnarray}

Using (\ref{PH0}), we obtain zeroth order Hamilton's equations
\begin{eqnarray}
\label{zerothHamil1}
&&\partial_tx_0 = y_0 {}\pi_{x0}, ~~~\partial_t{}\pi_{x0} = - x_0{}^3 \\
\label{zerothHamil2}
&&\partial_tx_0 = y_0 {}\pi_{y0}, ~~~\partial_t{}\pi_{y0} = - \frac{1}{2} \left(({}\pi_{x0}{}^2 + {}\pi_{y0}{}^2\right) - y_0{}^3
\end{eqnarray}

Using (\ref{PH2}), first order Hamilton's equations  are

\begin{eqnarray}
\label{firstHamil1}
&& \frac{\partial\mathcal{H}_2}{\partial {}\pi_{x1}} = 0~~~\Rightarrow {}\pi_{x0} y_1 + {}\pi_{x1} y_0 = 0\\
\label{firstHamil2}
&& \partial_ty_1 = \frac{\partial\mathcal{H}_2}{\partial{}\pi_{y1}} = {}\pi_{y0} y_1 + {}\pi_{y1} y_0 \\
\label{firstHamil3}
&& \partial_t{}\pi_{y1} = - \left( {}\pi_{x0} {}\pi_{x1} + {}\pi_{y0} {}\pi_{y1} \right) - 3 y_0{}^2 y_1
\end{eqnarray}

Equation (\ref{firstHamil1}) gives explicit expression for ${}\pi_{x1}$
and leads to identical expression as in (\ref{pmomentumx}). It can
easily be verified that, zeroth order equations (\ref{zerothHamil1})
and (\ref{zerothHamil2}) are identical to equations (\ref{zerothx})
and (\ref{zerothy}), respectively, where (\ref{firstHamil1}),
(\ref{firstHamil2}) and (\ref{firstHamil3}) lead to the equivalent
equation of motion of $y_1$ (\ref{firstoy1}).

To compare the the above expression with that from the Lagrangian
formulation, we rewrite the above expressions using (\ref{pmomentumx})
and (\ref{pmomentumy}). We get
\begin{eqnarray}
&&\mathcal{H}_0 = \frac{1}{2}\, (\partial_t{x_0}\, ){}^{2} y_0{}^{-1} + \frac{1}{2}\, (\partial_t{y_0}\, ){}^{2} y_0{}^{-1} + \frac{1}{4}\, x_0{}^{4} + \frac{1}{4}\, y_0{}^{4} \\
&&\mathcal{H}_1 = - \frac{1}{2}\, (\partial_t{x_0}\, ){}^{2} y_0{}^{(-2)} y_1 - \frac{1}{2}\, (\partial_t{y_0}\, ){}^{2} y_0{}^{(-2)} y_1 + \partial_t{y_0}\,  \partial_t{y_1}\,  y_0{}^{-1} +\nonumber \\
&&~~~~~~~~ y_0{}^{3} y_1 \\
&&\mathcal{H}_2 = - \frac{1}{2}\, (\partial_t{x_0}\, ){}^{2} y_0{}^{(-3)} y_1{}^{2} - \frac{1}{2}\, (\partial_t{y_0}\, ){}^{2} y_0{}^{(-3)} y_1{}^{2} + \frac{1}{2}\, (\partial_t{y_1}\, ){}^{2} y_0{}^{-1} + \nonumber \\
&&~~~~~~~~\frac{3}{2}\, y_0{}^{2} y_1{}^{2} \\
&&\mathcal{H}_3 = \frac{1}{2}\, (\partial_t{x_0}\, ){}^{2} y_0{}^{(-4)} y_1{}^{3} + \frac{1}{2}\, (\partial_t{y_0}\, ){}^{2} y_0{}^{(-4)} y_1{}^{3} - \partial_t{y_0}\,  \partial_t{y_1}\,  y_0{}^{(-3)} y_1{}^{2} +\nonumber \\
&&~~~~~~~~ \frac{1}{2}\, (\partial_t{y_1}\, ){}^{2} y_0{}^{(-2)} y_1 + y_1{}^{3} y_0 \\
&&\mathcal{H}_4 = \frac{1}{4} y_1{}^4
\end{eqnarray}

It is important to note that, only the third order perturbed
Hamiltonian is negative of the third order Lagrangian, i.e.,
\begin{equation}
\mathcal{H}_3 = - \mathcal{L}_3.
\end{equation}

Explicit forms of Interaction Hamiltonians can be obtained using
perturbed parts of the Lagrangian\cite{Huang:2006eha} and using
(\ref{secondL}), (\ref{thirdL}) and (\ref{fourthL}), it can be
verified that, both approaches lead to the identical results. In this
approach, to obtain $n^{th}$ order interaction Hamiltonian, the
Lagrangian is expanded up to $n^{th}$ order perturbation and by
varying the Lagrangian, the momentum corresponding to the perturbed
quantity is obtained as a non-linear combination of time derivative
of the field. Using perturbation techniques, this relation is inverted
and the time derivative of the field is written in terms of non-linear
combination of corresponding momentum. Perturbed Hamiltonian
corresponding to the perturbed parts of the Lagrangian is obtained by
using the conventional definition of Hamiltonian as $\mathcal{H} =
\pi\,\dot{\varphi} - \mathcal{L}$ and $\dot{\varphi}$ is replaced by
the above relation. Once the perturbed Hamiltonian is obtained in
terms of field variable and conjugate momentum, to calculate
correlation functions, the momentum in the Hamiltonian is replaced in
terms of time derivative of the field and in this time, the linear
relation of momentum and time derivative of the field is used. The
above procedure is rather cumbersome and involves series of
approximations. Since we start with the general Hamiltonian, our
approach is very straightforward and efficient.

As mentioned earlier, while we focus on first order perturbations, our
approach can easily be extended to any higher order perturbation,
e.g., to obtain second order equations of motion of field variables,
we have to consider up to second order field perturbation and its
corresponding momentum up to second order and calculate the fourth
order perturbed Hamiltonian. Since, we already
have obtained zeroth and first order equations, second order perturbed
field equations are obtained by varying fourth order perturbed
Hamiltonian with respect to second order perturbed variables and their
corresponding momenta.

\section{Canonical scalar field}\label{EoMLC}
In order to show the advantages of the Hamiltonian formulation, we
first focus on canonical scalar field. The action (\ref{AA}) for canonical
scalar field in the ADM formulation is
\begin{eqnarray}
\label{ActionCanonical}
&& \mathcal{S}_C = \int d^4 x \Big[N  \frac{\gamma^{\frac{1}{2}}}{2 \kappa}
\left(^{(3)}R + K_{i j} K^{i j} - K^2 \right) + \frac{1}{2}\, N{}^{-1} \gamma{}^{\frac{1}{2}} ({\partial}_{0}{\varphi}\, ){}^{2} -{N}^{i} {\partial}_{0}{\varphi}\,  {\partial}_{i}{\varphi}\,  N{}^{-1} \gamma{}^{\frac{1}{2}} \nonumber \\
&&~~~~~~~~~~  - \frac{1}{2}\, N {\gamma}^{i j} {\partial}_{i}{\varphi}\,  {\partial}_{j}{\varphi}\,  \gamma{}^{\frac{1}{2}} + \frac{1}{2}\, {N}^{i} {N}^{j} {\partial}_{i}{\varphi}\,  {\partial}_{j}{\varphi}\,  N{}^{-1} \gamma{}^{\frac{1}{2}} - N V \gamma{}^{\frac{1}{2}}\Big].
\end{eqnarray}

To obtain the equations of motion of all variables, one can simply use
the Einstein's equation or one can directly vary the action with
respect to field variables. 0-0 and 0-i components of the Einstein's
equations represent the equations of motion of $g_{0 0}$ and $g_{0
  i}$, which is $N$ and $N^{i}$ respectively. Hence, the above two
equations are identical to Hamiltonian and Momentum constraints
respectively and i-j component of the Einstein's equations and
conservation of Energy-Momentum tensor lead to equation of motion of
3-metric and matter field, respectively. In the rest of this
section, we will use the definitions (\ref{00pmetric}),
(\ref{0ipmetric}), (\ref{3metric}) and (\ref{field}) in the the above
equations to obtain perturbed equations of motion of
gravitational field variables at all order. In Appendix \ref{PLeCSF},
  we derive zeroth and first order perturbed field equations of
  canonical scalar field and in the following subsection, we apply the
  procedure discussed in section \ref{simple model} to canonical scalar field
  model to obtain consistent equations of motion as well as
  interaction Hamiltonian using Hamiltonian formulation.

\subsection{Hamitonian formulation}\label{EoMHC}
Conjugate momenta of all field variables $\gamma_{i j},\, \varphi,\,
N$ and $N^{i}$ are defined as
\begin{equation}
\pi^{i j} \equiv \frac{\delta L}{ \delta \dot{\gamma_{i j}}}, ~~\pi_\varphi \equiv \frac{\delta L}{ \delta \dot{\varphi}},~~~\pi_N \equiv \frac{\delta L}{ \delta \dot{N}}, ~~~\pi_{i}  \equiv \frac{\delta L}{ \delta \dot{N^{i}}}.
\end{equation}
Using the action (\ref{ActionCanonical}), conjugate momenta are given by
\begin{eqnarray}
\label{pimnC}
&& {\pi}^{i j} = \frac{1}{2}\, \kappa{}^{-1} \gamma{}^{\frac{1}{2}} ({\gamma}^{i j} {\gamma}^{k l} - {\gamma}^{i k} {\gamma}^{j l}) {K}_{k l} \\
\label{pipC}
&& \pi_\varphi  = N{}^{-1} \gamma{}^{\frac{1}{2}} \varphi^\prime\,  - {N}^{i} {\partial}_{i}{\varphi}\,  N{}^{-1} \gamma{}^{\frac{1}{2}} \\
\label{pinpini}
&& \Phi^N \equiv \pi_N = 0, ~~~\Phi^{N^{i}}_i \equiv \pi_{i} = 0.
\end{eqnarray}

From equation (\ref{pinpini}), it is apparent that, for canonical
scalar field, Lapse function $N$ and shift vector $N^{i}$ are
constraints and behave like Lagrange multipliers and it gives
4-primary constraint relations. Inverse relations of equations
(\ref{pimnC}) and (\ref{pipC}) are
\begin{eqnarray}
&& \gamma_{m n}^\prime = {\gamma}_{n k} {{N}^{k}{}_{|m}}\,  + {\gamma}_{m k} {{N}^{k}{}_{|n}}\,   - 2\, N {K}_{m n}, ~~ {K}_{i j} \equiv \kappa \gamma{}^{-\frac{1}{2}} ({\gamma}_{i j} {\gamma}_{k l} - 2\, {\gamma}_{i k} {\gamma}_{j l}) {\pi}^{k l} \\
&& \varphi^\prime = N \pi_\varphi \gamma{}^{-\frac{1}{2}} + {N}^{i} {\partial}_{i}{\varphi}
\end{eqnarray}

Using the above definitions of canonical momenta and
(\ref{ActionCanonical}), we get the Hamiltonian density of the system
\begin{eqnarray}
\label{HamiltonianCanonical}
&& \mathcal{H}_C = \pi^{i j} \gamma_{i j}^\prime + \pi_\varphi \varphi^\prime - L \nonumber\\
&&~~~~={\gamma}_{j k} {\partial}_{i}{{N}^{k}}\,  {\pi}^{i j} + {N}^{k} {\partial}_{k}{{\gamma}_{i j}}\,  {\pi}^{i j} + {\gamma}_{i k} {\partial}_{j}{{N}^{k}}\,  {\pi}^{i j} - N {\gamma}_{i j} {\gamma}_{k l} \kappa {\pi}^{k l} {\pi}^{i j} \gamma{}^{-\frac{1}{2}} + \nonumber \\
&&~~~~~~2\, N {\gamma}_{i k}\, {\gamma}_{j l}\, \kappa {\pi}^{k l} {\pi}^{i j} \gamma{}^{-\frac{1}{2}} + \frac{1}{2}\, N \pi_\varphi{}^{2} \gamma{}^{-\frac{1}{2}} + {N}^{i} \pi_\varphi {\partial}_{i}{\varphi}\,  - \frac{1}{2}\, N ^{(3)}R \gamma{}^{\frac{1}{2}} \kappa{}^{-1} +\nonumber \\
&&~~~~~~ \frac{1}{2}\, N {\gamma}^{i j} {\partial}_{i}{\varphi}\,  {\partial}_{j}{\varphi}\,  \gamma{}^{\frac{1}{2}} + N V \gamma{}^{\frac{1}{2}}.
\end{eqnarray}

Since the action (\ref{ActionCanonical}) has
diffeomorphism-invariance, Hamiltonian (\ref{HamiltonianCanonical})
vanishes identically, i.e., $$\mathcal{H}_C = 0.$$ Evolution of primary
constraints vanishes weakly and gives rise to four secondary
constraints: one Hamiltonian constraint due to $\mathcal{H}_N \equiv
\frac{d\Phi^N}{dt} =\{\pi_N, \mathcal{H}_C\}\equiv -\frac{\delta \mathcal{H}_C}{\delta N}
\approx 0$ and three Momentum constraints due to $\mathcal{H}_i \equiv
\frac{d\Phi^{N^{i}}_i}{dt} = \{\pi_i, \mathcal{H}_C\} \equiv - \frac{\delta
  \mathcal{H}_C}{\delta N^{i}} \approx 0$ .
\begin{eqnarray}
\label{HamConsCan}
&& \mathcal{H}_N \equiv  N \kappa \gamma{}^{-\frac{1}{2}} (2 {\gamma}_{i k}\, {\gamma}_{j l} - {\gamma}_{i j} {\gamma}_{k l}) \, \kappa {\pi}^{k l} {\pi}^{i j}  + \frac{1}{2}\, N \pi_\varphi{}^{2} \gamma{}^{-\frac{1}{2}} \nonumber \\
&& ~~~~~~ - \frac{1}{2}\, N ^{(3)}R \gamma{}^{\frac{1}{2}} \kappa{}^{-1} + \frac{1}{2}\, N {\gamma}^{i j} {\partial}_{i}{\varphi}\,  {\partial}_{j}{\varphi}\,  \gamma{}^{\frac{1}{2}} + N V \gamma{}^{\frac{1}{2}} \approx 0 \\
\label{MomConsCan}
&&\mathcal{H}_i \equiv  - 2 \partial_{k}{\gamma_{i j} \pi^{j k}} + \pi^{j k} \partial_{i}{\gamma_{j k}} + \pi_\varphi {\partial}_{i}{\varphi} \approx 0 .
\end{eqnarray}

Hamiltonian density can be written in terms Hamiltonian and Momentum
constraint as
\begin{equation}
\mathcal{H}_C = N \mathcal{H}_N + N^{i} \mathcal{H}_i \approx 0.
\end{equation}

\subsubsection{Zeroth order Hamilton's equations}
Using $\gamma_{i j} = a^2 \delta_{i j}$ and all background quantities
being independent of spatial coordinates, Hamiltonian density
(\ref{HamiltonianCanonical}) becomes
\begin{equation}
\label{HCB1}
\mathcal{H}_0^C =  2\, N_0\, a\, \kappa \,({\delta}_{i k} {\delta}_{j l}  - {\delta}_{i j} {\delta}_{k l}) \pi_0{}^{i j} {{}\pi_0{}}^{k l} + \frac{1}{2}\, N_0 {}\pi_{\varphi0}{}^{2} a{}^{(-3)} + N_0 V a{}^{3} \approx 0.
\end{equation}
At zeroth order, conjugate momentum of $a$, $\pi_a \equiv \frac{\delta
  L}{\delta \dot{a}}$ is directly related with ${}\pi_0{}^{i j}$ by the
simple relation ${}\pi_0{}^{i j} = \frac{1}{6 a} \pi_a \delta^{i j},~\pi_a
= 2 a \delta_{i j} {}\pi_0{}^{i j}$. The zeroth order Hamiltonian
(\ref{HCB1}), in terms of $\pi_a$ takes the simple form

\begin{equation}
\label{HCBF}
\mathcal{H}_0^C = N_0 \left[- \frac{1}{12}\,  \kappa \pi_a{}^{2} a{}^{-1} + \frac{1}{2}\,  {}\pi_{\varphi0}{}^{2} a{}^{(-3)} +  V a{}^{3}\right] \equiv N_0\, ^{(0)}\mathcal{H}_N.
\end{equation}

The terms inside the bracket in the right hand side is the Hamiltonian
constraint. At zeroth order it is independent of $N_0$. So, as we have
mentioned earlier, $N_0$ cannot be determined uniquely and we can
choose it arbitrarily. In this work, we use comoving coordinate, $N_0
= a$.

Varying the Hamiltonian (\ref{HCBF}) with respect to the momenta, we get
\begin{eqnarray}
&& a^\prime = - 1/6 N_0 \kappa \pi_a a^{-1} \\
&&\Rightarrow \pi_a = - 6 a a^\prime N_0^{-1} \kappa^{-1}~~~\Rightarrow {}\pi_0{}^{i j} = -N_0{}^{-1} \kappa{}^{-1} a^{\prime}\,  {\delta}^{i j} \\
&&\varphi_0^\prime = N_0 {}\pi_{\varphi0} a{}^{(-3)}
\end{eqnarray}

Hamiltonian constraint in conformal or comoving coordinate in terms of
field derivatives is
\begin{eqnarray}
\label{HConsCan0}
&&\mathcal{H}_{N0} \equiv - \frac{1}{12}\, \kappa \pi_a{}^{2} a{}^{-1} + \frac{1}{2}\, {}\pi_	{\varphi0}{}^{2} a{}^{(-3)} + V a{}^{3} = 0\\
\label{friedmann00}
&&\Rightarrow - 3\, \kappa{}^{-1} H^2 + \frac{1}{2}\,  \varphi_0^{\prime}\, {}^{2} + V\, a^2 = 0, \quad \mbox{where} \quad H \equiv \frac{a^\prime}{a}
\end{eqnarray}

Variation of the Hamiltonian with respect to the field variables and
relating with the time derivatives of the momenta lead to the
dynamical equation of motion of the field variables. Hence, equation
of motion of $a$ in comoving coordinate in terns of field derivatives
becomes
\begin{eqnarray}
\label{EoaHC0}
&&\pi_a^\prime\,  + \frac{\delta \mathcal{H}^C_0}{\delta a} = 0 \nonumber \\
&&\Rightarrow \pi_a^\prime\,  + \frac{1}{12} \, N_0 \kappa \pi_a{}^{2} a{}^{(-2)} - \frac{3}{2}\, N_0 {}\pi_{\varphi0}{}^{2} a{}^{(-4)} + 3\, N_0 V a{}^{2} = 0 \\
\label{eomah}
&&\Rightarrow 3\, \kappa{}^{-1} H^2 - 6\, \frac{a^{\prime \prime}}{a}\,  \kappa{}^{-1} - \frac{3}{2}\, \varphi_0^{\prime}\, {}^{2} a + 3\, V a{}^{2} = 0.
\end{eqnarray}

Similarly, the equation of motion of $\varphi_0$ takes the form
\begin{eqnarray}
\label{EovarphiHC0}
&&~~~~{}\pi_{\varphi0}\,{}^\prime  + N_0 V_\varphi\,  a{}^{3} = 0\\
\label{eomphih}
&& \Rightarrow \varphi_0^{\prime \prime} + 2 H\, \varphi_0^{\prime}+ V_\varphi\,  a{}^{2} = 0.
\end{eqnarray}
The three equations (\ref{friedmann00}), (\ref{eomah}) and
(\ref{eomphih}) are, as expected, identical to the equations
(\ref{B00EE}), (\ref{BijEE}) and (\ref{BEoMS}) respectively.

\subsubsection{First order perturbed Hamilton's equation}

As mentioned earlier, we consider flat-slicing gauge, hence there is
no perturbation in the 3-metric, i.e., $\delta g_{i j} = 0$. As we
pointed out in the simple model, while $x$ is treated as unperturbed,
$\pi_x$ is non-zero. In the flat-slicing gauge, there is no
perturbation in the 3-metric, however, canonical conjugate momentum
corresponding to 3-metric will have non-zero perturbed
contributions. This becomes transparent if we perturb (\ref{pimnC}),
i.e., $$\delta {\pi}^{m n} = \frac{1}{2}\, \kappa{}^{-1}
\gamma^{\frac{1}{2}} ({\gamma}^{m n} {\gamma}^{k l} - {\gamma}^{m k}
{\gamma}^{n l}) \delta {K}_{k l}.$$ Hence, perturbed part of $K_{i
  j}$, i.e., $\delta K_{i j}$ is not zero and it contributes to the
perturbed part of $\pi^{i j}$, i.e., $\delta \pi^{i j}$.

We can separate unperturbed and perturbed parts of field variables and
their corresponding momenta as
\begin{eqnarray}
\label{perturbfm1}
&& N = N_0 + \epsilon N_1, ~~~N^{i} = \epsilon N_1^i,~~~\varphi = \varphi_0 + \epsilon \varphi_1 \\
\label{perturbfm2}
&& \pi^{i j} = {}\pi_0{}^{i j} + \epsilon \pi_1^{i j}, ~~~\pi_\varphi = {}\pi_{\varphi0} + \epsilon \pi_{\varphi1}
\end{eqnarray}

Comparing (\ref{00pmetric}) and (\ref{0ipmetric}) with $N_1$ and $N_1^{i}$, we obtain 
\begin{equation}
\label{ncomp}
N_1 = a \,\phi_1,~~~~~~N_1^{i} = \delta^{i j}\, \partial_{i j}B_1
\end{equation}

The second order Hamiltonian density can be obtained by substituting
(\ref{perturbfm1}) and (\ref{perturbfm2}) in
(\ref{HamiltonianCanonical})
\begin{eqnarray}
\label{PHamSecCan}
&&\mathcal{H}^C_2 = {\delta}_{i j} {\partial}_{k}{{N_1}^{j}}\,  {\pi_1}^{i k} a{}^{2} + {\delta}_{i j} {\partial}_{k}{{N_1}^{j}}\,  {\pi_1}^{i k} a{}^{2} - N_0 {\delta}_{i j} {\delta}_{k l} \kappa {\pi_1}^{i j} {\pi_1}^{k l} a -  \nonumber \\
&&~~~~~~~~2\, N_1 {\delta}_{i j} {\delta}_{k l} \kappa \pi_0{}^{i j} {\pi_1}^{k l} a +2\, N_0 {\delta}_{i j} {\delta}_{k l} \kappa {\pi_1}^{i k} {\pi_1}^{j l} a + 4\, N_1 {\delta}_{i j} {\delta}_{k l} \kappa {{}\pi_0{}}^{i k} {\pi_1}^{j l} a +  \nonumber \\
&&~~~~~~~~\frac{1}{2}\, N_0 \pi_{\varphi1}{}^{2} a{}^{(-3)} + N_1 {}\pi_{\varphi0} \pi_{\varphi1} a{}^{(-3)} +{N_1}^{i} {}\pi_{\varphi0} {\partial}_{i}{\varphi_1}\,  + \nonumber \\
&&~~~~~~~~\frac{1}{2}\, N_0 {\delta}^{i j} {\partial}_{i}{\varphi_1}\,  {\partial}_{j}{\varphi_1}\,  a + N_1 V_\varphi\,  a{}^{3} \varphi_1 + \frac{1}{2}\, N_0 V_{\varphi \varphi}\,  \varphi_1{}^{2} a{}^{3}.
\end{eqnarray}

Since there is no perturbation in the 3-metric $\gamma_{i j}$,
variation with respect to $\pi_1^{i j}$ gives rise to a separate
constraint equation from which we can extract the derived value of
$\pi_1^{i j}$.
\begin{eqnarray}
&& ~~~\frac{\delta \mathcal{H}_2^C}{\delta \pi_1^{i j}} = 0 \\
&& \Rightarrow {\delta}_{n j} {\partial}_{m}{{N_1}^{j}}\,  a{}^{2} + {\delta}_{m j} {\partial}_{n}{{N_1}^{j}}\,  a{}^{2} - 2\, N_0 {\delta}_{m n} {\delta}_{k l} \kappa {\pi_1}^{k l} a + 4\, N_0 {\delta}_{m k} {\delta}_{n l} \kappa {\pi_1}^{k l} a - \nonumber \\
&& ~~~2\, N_1 {\delta}_{m n} {\delta}_{k l} \kappa {{}\pi_0{}}^{k l} a + 4\, N_1 {\delta}_{m k} {\delta}_{n l} \kappa {{}\pi_0{}}^{k l} a = 0.
\end{eqnarray}

Multiplying above expression with $(\delta^{m n} \delta^{i j} -
\delta^{m i} \delta^{n j})$ gives
\begin{eqnarray}
\label{pi1ij}
&&\pi_1^{i j} = \frac{1}{2}\, N_0{}^{-1} \kappa{}^{-1} a {\delta}^{i j} {\partial}_{k}{{N_1}^{k}}\,  - \frac{1}{4}\, N_0{}^{-1} \kappa{}^{-1} a {\delta}^{k i} {\partial}_{k}{{N_1}^{j}}\,  - \frac{1}{4}\, N_0{}^{-1} \kappa{}^{-1} a {\delta}^{k j} {\partial}_{k}{{N_1}^{i}}\,  \nonumber \\
&&~~~~~~~~- N_0{}^{-1} N_1 \pi_0{}^{i j}.
\end{eqnarray}

The relation of time derivative of perturbed matter field $\varphi_1$
and conjugate momentum of $\varphi_1$, $(\pi_{\varphi1})$ are
\begin{eqnarray}
&&~~~\varphi_1^\prime = \frac{\delta \mathcal{H}_2^C}{\delta \pi_{\varphi1}} \nonumber \\
&&\Rightarrow \varphi_1^\prime =  N_0 \pi_{\varphi1} a{}^{(-3)} +  N_1 {}\pi_{\varphi0} a{}^{(-3)}.
\end{eqnarray}

In the conformal coordinate, the perturbed Hamiltonian constraint is
obtained by varying the second order perturbed Hamiltonian density
(\ref{PHamSecCan}) with respect to perturbed Lapse function
$N_1$. Using above relations for the momenta with the time derivatives of
the field variables and (\ref{ncomp}), the perturbed Hamiltonian
constraint becomes
\begin{eqnarray}
&& ~~~~~~~\frac{\delta \mathcal{H}_2^C}{\delta N_1} = 0 \\
&& \Rightarrow 2\, {\delta}^{i j} H\,  {\partial}_{i j}{B_1}\,  \kappa{}^{-1} + 6\, \phi_1 \kappa{}^{-1} H^2 + \varphi_0^{\prime}\,  \varphi_1^\prime\,   -  \nonumber \\
\label{HConsH1}
&&~~~~~~~~\phi_1 \varphi_0^{\prime}\, {}^{2}  +V_\varphi\,  a{}^{2} \varphi_1 = 0.
\end{eqnarray}

Similarly, Momentum constraint is given by
\begin{eqnarray}
&& ~~~M_{i} \equiv \frac{\delta \mathcal{H}_2^C}{\delta N_1^{i}} = 0 \\
\label{MConsH1}
&&\Rightarrow \varphi_0^{\prime}\,  {\partial}_{i}{\varphi_1}\,   - 2\, H\,  {\partial}_{i}{\phi_1}\,  \kappa{}^{-1}  = 0.
\end{eqnarray}
and the equation of motion of the perturbed scalar field $\varphi_1$ becomes
\begin{eqnarray}
  && ~~~{\pi_{\varphi1}{}^\prime} + \frac{\delta \mathcal{H}_2^C}{\delta \varphi_1} = 0 \\
  && \Rightarrow  \varphi_1^{\prime \prime}\, + 2H\, \varphi_1^{\prime}\,    - {\phi_1^\prime}\,  \varphi_0^{\prime}\,  + 2\, V_\varphi \,  \phi_1 a{}^{2} - {\delta}^{i j} \varphi_0^{\prime}\,  {\partial}_{i j}{B_1}\,    \nonumber \\
\label{EConsH1}
&&~~~~~~~~- {\delta}^{i j} {\partial}_{i j}{\varphi_1}\, + V_{\varphi \varphi}\,  a{}^{2} \varphi_1 = 0.
\end{eqnarray}

Equation (\ref{HConsH1}), (\ref{MConsH1}) and (\ref{EConsH1}) are
identical to the equations (\ref{100EE}), (\ref{10iEE}) and
(\ref{1EoMS}), respectively.

This is a very important result. Unlike Lagrangian formalism, choosing
a gauge in Hamiltonian formalism is not trivial.  But using the above
simple mechanism, it is possible to construct the perturbed
Hamiltonian and its equations of motion and can now be treated in the
same manner as Lagrangian formalism. It can, even, be extended to any
order of perturbation, e.g., to obtain second order equations of
motion of field variables, we have to extend the Hamiltonian at fourth
order perturbation in terms of second order field variables and the
second order momenta and vary the Hamiltonian with respect to second
order variables and its conjugate momenta. At second order, in
flat-slicing gauge, we will again obtain a constraint equation
$\frac{\partial \mathcal{H}_4}{\partial \pi_2^{i j}} = 0$ that will give the
expression of $\pi_2^{i j}$. 

Our proposed mechanism works for any other arbitrary
  gauge also. In Appendix \ref{udgcsf}, we obtain all consistent
  perturbed and unperturbed Hamilton's equations for Canonical scalar
  field in uniform density gauge. This mechanism can be applied to
  any generalized first order derivative theory like non-canonical
  scalar field.

\subsubsection{Single variable-Momentum Hamiltonian}
Since the second order Hamiltonian holds the dynamics of first order
perturbed variables, we can obtain a single variable-single
momentum effective Hamiltonian of the background and constraint 
equations, and replacing background momenta in
terms of time derivatives of the fields. Substituting $\pi_1^{i j}$
using (\ref{pi1ij}) and ${}\pi_0{}^{i j}$ and ${}\pi_{\varphi0}$ in the
second order Hamiltonian and using (\ref{ncomp}), we get

\begin{eqnarray}
  && \mathcal{H}_2^C = \frac{1}{2}\, {\delta}^{i j} {\delta}^{k l} {\partial}_{i j}{B_1}\,  {\partial}_{k l}{B_1}\,  \kappa{}^{-1} a{}^{2} - \frac{1}{2} {\delta}^{i j} {\delta}^{k l} {\partial}_{i k}{B_1}\,  {\partial}_{j l}{B_1}\,  \kappa{}^{-1} a{}^{2} + 2\, {\delta}^{i j} H\,  {\partial}_{i j}{B_1}\,  \phi_1 \kappa{}^{-1} a^2 \nonumber \\
  &&~~~~~~+3\, \kappa{}^{-1} H^{2} \phi_1{}^{2} a^2 + \frac{1}{2}\, \pi_{\varphi1}{}^{2} a{}^{(-2)} + \pi_{\varphi1} \varphi_0^{\prime}\,  \phi_1 + {\delta}^{i j} \varphi_0^{\prime}\,  {\partial}_{i}{B_1}\,  {\partial}_{j}{\varphi_1}\,  a{}^{2} \nonumber \\
  &&~~~~~~+ \frac{1}{2}\, {\delta}^{i j} {\partial}_{i}{\varphi_1}\,  {\partial}_{j}{\varphi_1}\,  a{}^{2} + V_{\varphi}\,  \phi_1 a{}^{4} \varphi_1 + \frac{1}{2}\, V_{\varphi \varphi}\,  \varphi_1{}^{2} a{}^{4}
\end{eqnarray}

First and second terms in the right hand side lead to boundary
term. Moreover, performing integration by-parts to the seventh term
and substituting \[\phi_1 = \frac{\kappa}{2 H}
\varphi_0^\prime\, \varphi_1\] in the Hamiltonian, we get
\begin{eqnarray}
&&\mathcal{H}_2^C =\frac{3}{4}\, \kappa \varphi_0^{\prime}\, {}^{2} \varphi_1{}^{2} a{}^{2} + \frac{1}{2}\, \pi_{\varphi1}{}^{2} a{}^{(-2)} + \frac{\kappa}{2 H}\, \pi_{\varphi1}  \varphi_0^{\prime}\, {}^{2}  \varphi_1  + \frac{1}{2}\, {\delta}^{i j} {\partial}_{i}{\varphi_1}\,  {\partial}_{j}{\varphi_1}\,  a{}^{2}  \nonumber \\
&&~~~~~~+\frac{\kappa}{2 H}\,  V_\varphi\,  \varphi_0^{\prime}\,   \varphi_1{}^{2} a{}^{4} + \frac{1}{2}\, V_{\varphi \varphi}\,  \varphi_1{}^{2} a{}^{4}
\end{eqnarray}

This is the single variable-single momentum Hamiltonian density
$(\varphi_1, ~\pi_{\varphi1})$.  Further, this can be expressed in
terms of Mukhanov-Sasaki variable-Momentum\footnote{ Mukhanov-Sasaki variable is also a gauge invariant quantity related to curvature perturbation. At first order, it is given by $$u_1 = \frac{a \varphi_0^\prime}{H} \mathcal{R}_1.$$} form. In flat-slicing gauge,
Mukhanov-Sasaki variable $u_1 = a \varphi_1$, hence, 
\[\pi_{\varphi1} = a \pi_{u1}.\]

Hence the Hamiltonian in terms of Mukhanov-Sasaki variable and its conjugate momenta takes the form\footnote{Lagrangian can be written as
\begin{equation}
\mathcal{L} = 	\pi_{\varphi} \varphi^\prime - \mathcal{H}, \nonumber
\end{equation}

Changing of variables $\varphi \rightarrow \frac{u}{a}, \pi_\varphi \rightarrow a \pi_u$ leaves the Lagrangian unchanged. Hence,
\begin{eqnarray}
\mathcal{L} &=& a \pi_u \left( \frac{u^\prime}{a} - \frac{a^\prime}{a^2} u  \right) - \mathcal{H} \nonumber \\
&=& \pi_u u^\prime - \frac{a^\prime}{a} u \pi_u - \mathcal{H} \nonumber
\end{eqnarray}

Hence the new Hamiltonian in terms of $u$ and $\pi_u$ takes the form
\begin{equation}
\mathcal{H}^u = \mathcal{H}(u) + \frac{a^\prime}{a} u \pi_u \nonumber
\end{equation}

}

\begin{eqnarray}
\mathcal{H}_2^u &=& \mathcal{H}^C_2(u) + \frac{a^\prime}{a}\, \pi_{u1} u_1 \nonumber \\ 
&=&\frac{3}{4}\, \kappa u_1{}^{2} \varphi_0^{\prime}\, {}^{2} + \frac{1}{2}\, \pi_{u1}{}^{2} + \frac{1}{2}\, \pi_{u1} u_1 \kappa \varphi_0^{\prime}\, {}^{2} H{}^{-1} + \frac{1}{2}\, {\delta}^{i j} {\partial}_{i}{u_1}\,  {\partial}_{j}{u_1}\,  \nonumber \\
&&+ \frac{1}{2}\, \kappa V_\varphi\,  \varphi_0^{\prime}\,  u_1{}^{2} H^{-1} a{}^{2} + \frac{1}{2}\, V_{\varphi\varphi}\,  u_1{}^{2} a{}^{2} + H\, \pi_{u1} u_1.
\end{eqnarray}

One can verify that the equation of motion of $u_1$ becomes $$u_1^{\prime \prime} - \nabla^2 u_1 - \frac{z^{\prime\prime}}{z} u_1 = 0, \quad \mbox{where} \quad z \equiv \frac{a \varphi_0^\prime}{H}.$$

\subsubsection{Interaction Hamiltonian for calculating higher order
  correlations} 
  
Expanding Hamiltonian
  (\ref{HamiltonianCanonical}) to third order, we obtain third order
  Interaction Hamiltonian whose explicit form in phase space is given
  in Appendix \ref{ThirdOrderInteractionCanonical}. Replacing the
  momenta in terms of terms of time derivatives of the fields,
  interaction Hamiltonian for canonical scalar field becomes
{
\begin{eqnarray}
&& \mathcal{H}_3^C \, = - \frac{1}{2}\, {\delta}^{i j} {\delta}^{k l} {\partial}_{i j}{B_1}\,  {\partial}_{k l}{B_1}\,  \phi_1 \kappa{}^{-1} a{}^{2} - 2\, {\delta}^{i j} H\,  {\partial}_{i j}{B_1}\,  \kappa{}^{-1} \phi_1{}^{2} a^2 - 
3\, \kappa{}^{-1} H^{2} \phi_1{}^{3} a^2 + \nonumber \\
&&\frac{1}{2}\, {\delta}^{i j} {\delta}^{k l} {\partial}_{i k}{B_1}\,  {\partial}_{j l}{B_1}\,  \phi_1 \kappa{}^{-1} a{}^{2} +  \frac{1}{2}\, \phi_1 {\varphi_1^\prime}\, {}^{2} a{}^{2} - {\varphi_0^\prime}\,  {\varphi_1^\prime}\,  \phi_1{}^{2} a{}^{2} + \frac{1}{2}\, {\varphi_0^\prime}\, {}^{2} \phi_1{}^{3} a{}^{2} + \nonumber \\
&&{\delta}^{i j} {\varphi_1^\prime}\,  {\partial}_{i}{B_1}\,  {\partial}_{j}{\varphi_1}\,  a{}^{2} - {\delta}^{i j} {\varphi_0^\prime}\,  {\partial}_{i}{B_1}\,  {\partial}_{j}{\varphi_1}\,  \phi_1 a{}^{2} + \frac{1}{2}\, {\delta}^{i j} {\partial}_{i}{\varphi_1}\,  {\partial}_{j}{\varphi_1}\,  \phi_1 a{}^{2} + \frac{1}{2}\, V_{\varphi\varphi}\,  \phi_1 \varphi_1{}^{2} a{}^{4} + \nonumber \\
&&\frac{1}{6}\, V_{\varphi\varphi\varphi}\,  \varphi_1{}^{3} a{}^{4}.
\end{eqnarray}}

It can be verified that $^{(3)}\mathcal{H}_C = - ^{(3)}\mathcal{L}_C$. Similarly, fourth order Interaction Hamiltonian for canonical scalar field takes the form
{
\begin{equation}
\mathcal{H}_4^C = \frac{1}{6}\, \phi_1 V_{\varphi \varphi \varphi}\,  \varphi_1{}^{3} a{}^{4} + \frac{1}{24}\,  V_{\varphi \varphi \varphi \varphi}\,  \varphi_1{}^{4} a{}^{4}
\end{equation}}
which is independent of kinetic part (time derivatives of fields) of
the field. This can be verified by looking at the Hamiltonian
(\ref{HamiltonianCanonical}). Terms containing momenta only contribute
up to third order Hamiltonian since $\gamma_{i j}$ is unperturbed.
Hence, fourth or higher order perturbed Hamiltonian is independent of
the kinetic part of the fields. {Furthermore, the higher order interaction Hamiltonian can be expressed as single variable form by using constrained equations (\ref{HConsH1}) and (\ref{MConsH1}).}

\section{Galilean single scalar field model}\label{galileanfield}
In the action (\ref{boxx}), $G(X, \varphi) \neq 0$ leads to Galilean
field where action contains second derivative terms of the field
variables. In this present work, to simplify our calculations, we take
a specific and simple form of the Galilean model with $P(X, \varphi) =
- V(\varphi)$ and $G(X, \varphi) = -2 X$, i.e.,
\begin{equation}
\label{action}
\mathcal{S}_G = \int d^4 x \sqrt{-g} \left[\frac{1}{2 \kappa} R -  g^{\mu \nu} \partial_{\mu}{\varphi} \partial_{\nu}{\varphi} \Box \varphi  - V(\varphi) \right] \equiv \int \mathcal{L}_G~d^4 x.
\end{equation}

Zeroth and first order perturbed Euler-Lagrange equations of motion are provided in Appendix \ref{galmod}.

\subsection{Hamiltonian formulation of the Galilean scalar field}

Hamiltonian formulation of Higher derivative fields is not unique and there
exist several ways~\cite{Ostro,Andrzejewski:2010kz,Deffayet:2015qwa,Nesterenko,Chen:2012au,Woodard:2015zca}
to rewrite the Hamiltonian as there are infinite ways to absorb the higher
derivative terms. For our case, one easy way is to let 
$S \equiv \Box \varphi$ and re-write the action (\ref{action}) as
\begin{equation}
\label{full action galilean}
\mathcal{S}_G = \int d^4 x \sqrt{-g} \left[\frac{1}{2 \kappa} R - \frac{1}{2} g^{\mu \nu} \partial_{\mu}{\varphi} \partial_{\nu}{\varphi} S  - V(\varphi) \right] + \int d^4 x ~\lambda \left(S - \Box \varphi\right).
\end{equation}
where $\lambda$ is the Langrange multiplier whose variation leads to 
$S = \Box \varphi$. Since $\Box \varphi$ appears linearly in the 
action, we can rewrite the action in terms of
first order derivatives of the fields by performing integration
by-parts as
\begin{eqnarray}
&& \mathcal{S}_G = \int d^4 x \sqrt{-g} \left[\frac{1}{2 \kappa} R - \frac{1}{2} g^{\mu \nu} \partial_{\mu}{\varphi} \partial_{\nu}{\varphi} \, S  - V(\varphi) \right] + \int d^4 x ~\lambda \left(S - \Box \varphi\right) \nonumber\\
&& ~~= ... + \int d^4x \left[\lambda S - \lambda~ g^{\mu \nu} \left(\partial_{\mu \nu}{\varphi} - \Gamma^{\alpha}_{\mu \nu} ~\partial_{\alpha}{\varphi} \right) \right] \nonumber \\
&& ~~= ... + \int d^4x \left[\lambda S + \lambda~ g^{\mu \nu}~\Gamma^{\alpha}_{\mu \nu} ~\partial_{\alpha}{\varphi} + g^{\mu \nu} \partial_{\mu}{\varphi} \partial_{\nu}\lambda + \lambda \, \partial_{\nu}{g^{\mu \nu}} \partial_{\mu}{\varphi}\right] + Boundary~term \nonumber \\
\label{full action}
&& ~~= \int d^4 x \sqrt{-g} \left[\frac{1}{2 \kappa} R - \frac{1}{2} g^{\mu \nu} \partial_{\mu}{\varphi} \partial_{\nu}{\varphi} \, S  - V(\varphi) \right] + \int d^4x [\, \lambda S + \lambda~ g^{\mu \nu}~\Gamma^{\alpha}_{\mu \nu} ~\partial_{\alpha}{\varphi} + \nonumber \\
&& ~~~~~~~~g^{\mu \nu} \partial_{\mu}{\varphi} \partial_{\nu}\lambda + \lambda \, \partial_{\nu}{g^{\mu \nu}} \partial_{\mu}{\varphi}\,]. 
\end{eqnarray}
Although the action, now, contains extra variables $\lambda $ and $S$,
action (\ref{action}) and (\ref{full action}) lead to equivalent
equations of motion. Since, the action (\ref{full action}) contains no
double derivative terms, we can construct the Hamiltonian of the
system.

Using the line element (\ref{line}), the action (\ref{full
  action}) can be decomposed and rewritten as

{\small
\begin{eqnarray}
\label{full lagrangian}
 &&\mathcal{S}_G \equiv \int \mathcal{L}_G\,d^4x = \int d^4 x \,(S \lambda - N V \gamma{}^{\frac{1}{2}} + {\gamma}^{i j} {\partial}_{i}{\lambda}\,  {\partial}_{j}{\varphi}\,  + \lambda {\partial}_{i}{{\gamma}^{i j}}\,  {\partial}_{j}{\varphi}\,  - {\lambda^\prime}\,  {\varphi^\prime}\,  N{}^{(-2)} + \frac{1}{2}\, N ^{(3)}R \gamma{}^{\frac{1}{2}} \kappa{}^{-1} + \nonumber \\
 &&~~~~{N}^{i} {\lambda^\prime}\,  {\partial}_{i}{\varphi}\,  N{}^{(-2)} + {N}^{i} {\varphi^\prime}\,  {\partial}_{i}{\lambda}\,  N{}^{(-2)} + S N{}^{-1} \gamma{}^{\frac{1}{2}} {\varphi^\prime}\, {}^{2} + \lambda {N^\prime}\,  {\varphi^\prime}\,  N{}^{(-3)} - {N}^{i} {N}^{j} {\partial}_{i}{\lambda}\,  {\partial}_{j}{\varphi}\,  N{}^{(-2)} - \nonumber \\
 &&~~~~{N}^{i} \lambda {N^\prime}\,  {\partial}_{i}{\varphi}\,  N{}^{(-3)} - {N}^{i} \lambda {\varphi^\prime}\,  {\partial}_{i}{N}\,  N{}^{(-3)} + {\gamma}^{i j} {\gamma}^{k l} \lambda {\partial}_{i}{{\gamma}_{j k}}\,  {\partial}_{l}{\varphi}\,  - \frac{1}{2}\, {\gamma}^{i j} {\gamma}^{k l} \lambda {\partial}_{i}{{\gamma}_{k l}}\,  {\partial}_{j}{\varphi}\,  + \nonumber \\
 &&~~~~\frac{1}{2}\, {\gamma}^{i j} \lambda {{\gamma}_{i j}}{}^\prime\,  {\varphi}^\prime\,  N{}^{(-2)} - {\gamma}^{i j} \lambda {\partial}_{i}{N}\,  {\partial}_{j}{\varphi}\,  N{}^{-1} - N S {\gamma}^{i j} {\partial}_{i}{\varphi}\,  {\partial}_{j}{\varphi}\,  \gamma{}^{\frac{1}{2}} + {N}^{i} {N}^{j} \lambda {\partial}_{i}{N}\,  {\partial}_{j}{\varphi}\,  N{}^{(-3)}%
 - \nonumber \\
 &&~~~~2\, {N}^{i} S {\varphi^\prime}\,  {\partial}_{i}{\varphi}\,  N{}^{-1} \gamma{}^{\frac{1}{2}} - \frac{1}{2}\, {N}^{i} {\gamma}^{j k} \lambda {{\gamma}_{j k}}{}^\prime\,  {\partial}_{i}{\varphi}\,  N{}^{(-2)} - \frac{1}{2}\, {N}^{i} {\gamma}^{j k} \lambda {\varphi^\prime}\,  {\partial}_{i}{{\gamma}_{j k}}\,  N{}^{(-2)} - \nonumber \\
 &&~~~~\frac{1}{2}\, {K}^{i j} {K}^{k l} N {\gamma}_{i j} {\gamma}_{k l} \gamma{}^{\frac{1}{2}} \kappa{}^{-1} + \frac{1}{2}\, {K}^{i j} {K}^{k l} N {\gamma}_{i k} {\gamma}_{j l} \gamma{}^{\frac{1}{2}} \kappa{}^{-1} + {N}^{i} {N}^{j} S {\partial}_{i}{\varphi}\,  {\partial}_{j}{\varphi}\,  N{}^{-1} \gamma{}^{\frac{1}{2}} +\nonumber \\
 &&~~~~ \frac{1}{2}\, {N}^{i} {N}^{j} {\gamma}^{k l} \lambda {\partial}_{i}{{\gamma}_{k l}}\,  {\partial}_{j}{\varphi}\,  N{}^{(-2)}).
\end{eqnarray}}

Momenta corresponding to the variables are defined as
\begin{eqnarray}
\label{dmomenta}
&&\pi^{i j} = \frac{\delta \mathcal{S}_G}{\delta \gamma_{i j}{}^\prime} ~~~~~~~~ \pi_N = \frac{\delta \mathcal{S}_G}{\delta N^\prime} ~~~~~~~~ \pi_{i} = \frac{\delta \mathcal{S}_G}{\delta N^{i}{}^\prime} \nonumber \\
&&\pi_\varphi = \frac{\delta \mathcal{S}_G}{\delta \varphi^\prime}~~~~~~~~\pi_\lambda = \frac{\delta \mathcal{S}_G}{\delta \lambda^\prime}~~~~~~~~ \pi_S = \frac{\delta \mathcal{S}_G}{\delta S^\prime} \nonumber ~~
\end{eqnarray}

Using the action (\ref{full lagrangian}), we obtain the following relations:
\begin{eqnarray}
\label{momentag}
&& \pi^{i j} = \frac{1}{2}\, \kappa{}^{-1} \gamma{}^{\frac{1}{2}} ({\gamma}^{m n} {\gamma}^{k l} - {\gamma}^{m k} {\gamma}^{n l}) {K}_{k l} + \frac{1}{2}\, {\gamma}^{m n} \lambda {\varphi^\prime}\,  N{}^{(-2)} - \frac{1}{2}\, {N}^{i} {\partial}_{i}{\varphi}\,  \lambda N{}^{(-2)} {\gamma}^{m n} ~~~~~~~~ \\
\label{momentaN}
&& \pi_\varphi =  - {\lambda^\prime}\,  N{}^{(-2)} + {N}^{i} {\partial}_{i}{\lambda}\,  N{}^{(-2)} + 2\, S N{}^{-1} \gamma{}^{\frac{1}{2}} {\varphi^\prime}\,  + \lambda {N^\prime}\,  N{}^{(-3)} - {N}^{i} \lambda {\partial}_{i}{N}\,  N{}^{(-3)}\nonumber \\
&& ~~~~~~~~+ \frac{1}{2}\, {\gamma}^{i j} \lambda {{\gamma}_{i j}{}^\prime}\,  N{}^{(-2)} - 2\, {N}^{i} S {\partial}_{i}{\varphi}\,  N{}^{-1} \gamma{}^{\frac{1}{2}} - \frac{1}{2}\, {N}^{i} {\gamma}^{j k} \lambda {\partial}_{i}{{\gamma}_{j k}}\,  N{}^{(-2)} \\
\label{momentaL}
&& \pi_\lambda = - {\varphi^\prime}\,  N{}^{(-2)} + {N}^{i} {\partial}_{i}{\varphi}\,  N{}^{(-2)} \\
&& \pi_N = {\varphi^\prime}\,  \lambda N{}^{(-3)} - {N}^{i} \lambda {\partial}_{i}{\varphi}\,  N{}^{(-3)} \nonumber \\
\label{momentaNN}
&& ~~~~~~~~ = - \lambda N{}^{-1} \pi_\lambda \\
&& \pi_{i} = 0 \\
&& \pi_S = 0.
\end{eqnarray}

Equations (\ref{momentag}), (\ref{momentaN}) and (\ref{momentaL}) are
invertible and $\partial_{0}{\gamma_{i j}}$, $\partial_{0}{\lambda}$
and $\partial_{0}{\varphi}$ can be written in terms of $\pi^{i j}$,
$\pi_\varphi$ and $\pi_\lambda$ as

\begin{eqnarray}
\label{g0}
&& \gamma_{m n}{}^\prime = {\gamma}_{n k} {N}^{k}{}_{|m}\,  + {\gamma}_{m k} {{N}^{k}{}_{|n}}\,  - 2\, N {K}_{m n} \\
\label{k0}
&& K_{i j} = \kappa \gamma{}^{-1/2} ({\gamma}_{i j} {\gamma}_{m n} - 2\, {\gamma}_{i m} {\gamma}_{j n}) {\pi}^{m n} + \frac{1}{2}\, {\gamma}_{i j} \lambda \pi_\lambda \\
\label{v0}
&& {\varphi}{}^\prime = N^{i} \partial_{i}{\varphi} - N^2 \pi_\lambda \\
&& {\lambda}{}^\prime = N^2 ( - \pi_\varphi + 2 S N^{-1} \gamma^{1/2} {\varphi^\prime} - 2 N^{i} S \partial_{i}{\varphi} N^{-1} \gamma^{1/2} 
 + N^{i} \partial_{i}{\lambda} N^{(-2)} + \lambda {N^\prime} N^{(-3)} \nonumber \\
 \label{l0}
 &&~~~~~~~~- N^{i} \lambda \partial_{i}{N} N^{(-3)} + \frac{1}{2} \gamma^{i j} \lambda {\gamma_{i j}^\prime} N^{(-2)} - \frac{1}{2} N^{i} \gamma^{j k} \lambda \partial_{i}{\gamma_{j k}} N^{(-2)}).
\end{eqnarray}

 Hence the Hamiltonian density is given by

\begin{equation}
 \mathcal{H}_G = \pi^{i j} {\gamma_{i j}^\prime} + \pi_\varphi {\varphi^\prime} + \pi_{\lambda} {\lambda^\prime} + \pi_N {N^\prime} - \mathcal{L}_G.
\end{equation}

Using the action (\ref{full lagrangian}) and (\ref{g0}), (\ref{k0}), (\ref{v0}) (\ref{l0}) and (\ref{momentaNN}), the Hamiltonian density becomes
{\small
\begin{eqnarray}
\label{fullH}
&&\mathcal{H}_G =  - S \lambda + N V \gamma{}^{\frac{1}{2}} + {N}^{i} \pi_\lambda {\partial}_{i}{\lambda}\,  + {N}^{i} \pi_\varphi {\partial}_{i}{\varphi}\,  - \pi_\lambda \pi_\varphi N{}^{2} + \pi_\lambda \lambda {\partial}_{i}{{N}^{i}}\,  + 2 {\gamma}_{i j} {\partial}_{k}{{N}^{i}}\,  {\pi}^{ j k} +\nonumber \\
&&~~~~ {\gamma}^{i j} \lambda {\partial}_{i}{N}\,  {\partial}_{j}{\varphi}\,  N{}^{-1} - {\gamma}^{i j} {\partial}_{i}{\lambda}\,  {\partial}_{j}{\varphi}\,  - \lambda {\partial}_{i}{{\gamma}^{i j}}\,  {\partial}_{j}{\varphi}\,  - \frac{1}{2}\, N ^{(3)}R \gamma{}^{\frac{1}{2}} \kappa{}^{-1} - S N{}^{3} \pi_\lambda{}^{2} \gamma{}^{\frac{1}{2}} - \nonumber \\
&&~~~~\frac{3}{4}\, N \kappa \pi_\lambda{}^{2} \gamma{}^{-\frac{1}{2}} \lambda{}^{2} - {N}^{i} \pi_\lambda \lambda {\partial}_{i}{N}\,  N{}^{-1} +  {N}^{i} {\partial}_{i}{{\gamma}_{l m}}\,  {\pi}^{l m} -   N^{i} {\partial}_{l}{{\gamma}_{i m}}\,  {\pi}^{l m} + \nonumber \\
&&~~~~ {N}^{i} {\partial}_{m}{{\gamma}_{i l}}\,  {\pi}^{l m} - {\gamma}^{i j} {\gamma}^{k l} \lambda {\partial}_{i}{{\gamma}_{j k}}\,  {\partial}_{l}{\varphi}\,  + \frac{1}{2}\, {\gamma}^{i j} {\gamma}^{k l} \lambda {\partial}_{i}{{\gamma}_{k l}}\,  {\partial}_{j}{\varphi}\,  + N S {\gamma}^{i j} {\partial}_{i}{\varphi}\,  {\partial}_{j}{\varphi}\,  \gamma{}^{\frac{1}{2}} - \nonumber \\
&&~~~~    N \pi_\lambda {\gamma}_{i j} \kappa \lambda {\pi}^{i j} \gamma{}^{-\frac{1}{2}}+ 2\, N {\gamma}_{i j} {\gamma}_{k l} \kappa {\pi}^{i k} {\pi}^{j l} \gamma{}^{-\frac{1}{2}} - N {\gamma}^{i j} {\gamma}^{k l} \kappa {\pi}^{i j} {\pi}^{k l} \gamma{}^{-\frac{1}{2}}.
\end{eqnarray}}

Since $\pi_N$ does not appear in the Hamiltonian, complete dynamics of
the fields are obtained from the Dirac Hamiltonian
\begin{equation}
\label{DHamiltonian}
\mathcal{H}_D = \mathcal{H}_G + \xi \left( \pi_N + \frac{\lambda}{N} \pi_\lambda \right).
\end{equation}

The above Dirac-Hamiltonian can be used to obtain the perturbed
equations of motion.

\subsubsection{Background equations} \label{GalBack}
At zeroth order, field variables are defined as
\begin{eqnarray}
&&N = N_0, ~~~~ N^{i} = 0,~~~~ \gamma_{i j} = a^2 \delta_{i j}, ~~~~S= S_0,~~~~\lambda=\lambda_0,~~~~\varphi = \varphi_0 \nonumber\\
&&\pi^{i j} = {}\pi_0{}^{i j}, ~~~~ \pi_N = \pi_{N0},~~~~\pi_\varphi= {}\pi_{\varphi0},~~~~\pi_\lambda = {}\pi_{\lambda0}.
\end{eqnarray}
Using the above relations, the Dirac-Hamiltonian (\ref{DHamiltonian})
takes the form

\begin{eqnarray}
\label{BackHD}
&&\mathcal{H}_{D0} = - S_0 \lambda_0 + N_0 V_0 a{}^{3} - {}\pi_{\lambda0} {}\pi_{\varphi0} N_0{}^{2} - S_0 N_0{}^{3} {}\pi_{\lambda0}{}^{2} a{}^{3} - \frac{3}{4}\, N_0 \kappa {}\pi_{\lambda0}{}^{2} \lambda_0{}^{2} a{}^{(-3)} - \nonumber\\
&&~~~~~~~~\frac{1}{2}\, N_0 \pi_a {}\pi_{\lambda0} \kappa \lambda_0 a{}^{(-2)} - \frac{1}{12}\, N_0 \kappa \pi_a{}^{2} a{}^{-1} + \xi_0 \left( \pi_{N0}  + {}\pi_{\lambda0} \lambda_0 N_0{}^{-1} \right).
\end{eqnarray}

Since the momentum corresponding to $S$ does not appear in the
Hamiltonian (\ref{BackHD}), variation of the Hamiltonian with
respect to $S$ leads to the first secondary constraint and it is given
by
\begin{equation}
\label{lamb0}
\lambda_0 = -N_0{}^{3} {}\pi_{\lambda0}{}^{2} a{}^{3}
\end{equation}
and varying the Dirac-Hamiltonian with respect to $\xi_0$ recovers the primary constraint
\begin{equation}
\label{pin0constraint}
\pi_{N0} = - \lambda_0 {}\pi_{\lambda0} N_0^{-1}.
\end{equation}

The Hamilton's equations which relate time variation of the field
variables with momenta, are given by
\begin{eqnarray}
\label{del0phi0}
&& {\varphi_0{}^\prime} = - {}\pi_{\lambda0} N_0^2 \\
\label{del0a0}
&& a^{\prime}\,  =  - \frac{1}{6}\, N_0 \kappa a{}^{-1} \pi_a - \frac{1}{2}\, N_0 {}\pi_{\lambda0} \kappa \lambda_0 a{}^{(-2)} \\
\label{del0n0}
&& {N_0{}^\prime} = \xi_0 \\
\label{del0lam0}
&& {\lambda_0{}^\prime} = - {}\pi_{\varphi0} N_0{}^{2} - 2\, S_0 N_0{}^{3} {}\pi_{\lambda0} a{}^{3} - \frac{3}{2}\, N_0 \kappa {}\pi_{\lambda0} \lambda_0{}^{2} a{}^{(-3)} - \nonumber \\
&&~~~~~~~~\frac{1}{2}\, N_0 \pi_a \kappa \lambda_0 a{}^{(-2)} + \lambda_0 N_0{}^{-1} \xi_0 ~~~~~~~~~~\\
\end{eqnarray}
which relate the momenta and time derivatives of the fields. The above
relations are invertible and all momenta can be written in terms of
the time derivatives of the field variables. Hamilton's equations
corresponding to the time variation of momenta are given by:
\begin{eqnarray}
\label{pia}
&& {\pi_a{}^\prime} = - 3\, N_0 V_0 a{}^{2} + 3\, S_0 N_0{}^{3} {}\pi_{\lambda0}{}^{2} a{}^{2} - \frac{9}{4}\, N_0 \kappa {}\pi_{\lambda0}{}^{2} \lambda_0{}^{2} a{}^{(-4)} - N_0 \pi_a {}\pi_{\lambda0} \kappa \lambda_0 a{}^{(-3)}\nonumber \\
&&~~~~~~~~ - \frac{1}{12}\, N_0 \kappa \pi_a{}^{2} a{}^{(-2)} \\
\label{pil}
&& {{}\pi_{\lambda0}{}^\prime} =   S_0 - 3\, {}\pi_{\lambda0} a^{\prime}\,  a{}^{-1} - {}\pi_{\lambda0} {\partial}_{0}{N_0}\,  N_0{}^{-1} \\
\label{pin}
&& {\pi_{N0}{}^\prime} = - V_0 a{}^{3} + 2\, {}\pi_{\lambda0} {}\pi_{\varphi0} N_0 + 3\, S_0 N_0{}^{2} {}\pi_{\lambda0}{}^{2} a{}^{3} + \frac{3}{4}\, \kappa {}\pi_{\lambda0}{}^{2} \lambda_0{}^{2} a{}^{(-3)} \nonumber \\
&&~~~~~~~~+ \frac{1}{2}\, \pi_a {}\pi_{\lambda0} \kappa \lambda_0 a{}^{(-2)} + \frac{1}{12}\, \kappa \pi_a{}^{2} a{}^{-1} + {}\pi_{\lambda0} \lambda_0 N_0{}^{(-2)} \xi_0 \\
\label{pip}
&& {{}\pi_{\varphi0}{}^\prime} = - N_0 V_{\varphi} a^3.
\end{eqnarray}
which, by using other background Hamilton's equations, lead to the
identical dynamical equations of motion of the field variables
obtained from action formulation.  Using background Hamilton's
equations in conformal time coordinate, equation (\ref{pil}) becomes
\begin{equation}
\label{s0}
S_0 = - 2H\, \varphi_0^{\prime}\,  a{}^{(-2)} - \varphi_0^{\prime \prime}\,  a{}^{(-2)}.
\end{equation}

Variation of action (\ref{full action galilean}) or (\ref{full
  action}) with respect to $\lambda$ lead to $S = \Box \varphi$. At
zeroth order, $\Box \varphi = - 2H\, \varphi_0^{\prime}\, 
a{}^{(-2)} - \varphi_0^{\prime \prime}\, a{}^{(-2)}$ implying that the
dynamical equation (\ref{s0}) obtained using Hamiltonian formulation
is consistent. Similarly, equation (\ref{pin}) leads to the zeroth
order Hamiltonian constraint of the Galilean scalar field model. In
conformal time, it is given by
\begin{equation}
\label{hc0g}
V_0 a{}^{2} - 6\, H\,  a{}^{(-2)} \varphi_0^{\prime}\, {}^{3} - 3\,  \kappa{}^{-1} H^2 = 0
\end{equation}

Equations (\ref{pia}) and (\ref{pip}) lead to the equation of motion
of $a$ and $\varphi_0$ respectively and are given by
\begin{eqnarray}
\label{eoma0g}
&&6\, H\,  \varphi_0^{\prime}\, {}^{3} a{}^{(-2)} - 6\, \varphi_0^{\prime \prime}\,  \varphi_0^{\prime}\, {}^{2} a{}^{-2} + 3\, \kappa{}^{-1} H^2 - 6\, a^{\prime \prime}\, a{}^{-1} \kappa{}^{-1} + 3\, V_0 a{}^{2}= 0 ~~~~~~~~\\
\label{eomphig}
&& \frac{1}{2}\,\varphi_0^{\prime}{}^2\,a^{\prime \prime}\,a^{-1}H{}^{-1} -  \frac{H}{2}\,\varphi_0^{\prime}{}^2 + \,\varphi_0^{\prime \prime}\varphi_0^\prime\,  - \frac{1}{12}\, V_{\varphi}\,   H{}^{-1} a{}^{4} = 0.
\end{eqnarray}

Equations (\ref{hc0g}), (\ref{eoma0g}) and (\ref{eomphig}) are
identical to the Lagrangian equations (\ref{hc0g1}), (\ref{eomag1}) and (\ref{eomphig1}),
respectively. Hence, at zeroth order, Hamiltonian formulation is
consistent with Lagrangian formulation.

\paragraph{\underline{Counting scalar degrees of freedom at zeroth order:}}
As one can see in (\ref{BackHD}), background phase space contains 10
variable ($a,\, N, \, \varphi_0, \, \lambda_0, \,
S_0$ and corresponding momenta). There are two primary
constrained equations:
\begin{eqnarray}
\label{pc01}
&& \Phi^1_p \equiv \pi_{S0} = 0 \\
\label{pc02}
&& \Phi_p^2 \equiv \pi_{N0} + \lambda_0 {}\pi_{\lambda0} N_0^{-1} = 0
\end{eqnarray}

Conservation of primary constraints gives rise to secondary constraints:
\begin{eqnarray}
\label{sc01}
\Phi_s^1 \equiv \{\Phi_p^1, \mathcal{H}_{D0}\} \approx 0 && 
\Rightarrow \lambda_0 + N_0{}^{3} {}\pi_{\lambda0}{}^{2} a{}^{3} \approx 0 \\
\label{sc02}
\Phi_s^2 \equiv \{\Phi_p^2, \mathcal{H}_{D0}\} \approx 0 
&& \Rightarrow - V_0 a^3 + {}\pi_{\lambda0} {}\pi_{\varphi0} N_0 + \frac{3}{4 a^3} \kappa {}\pi_{\lambda0}{}^2 \lambda_0{}^2  + 
\frac{1}{12 \, a} \kappa \pi_a^2 \nonumber \\
&&~~~~+ \frac{\kappa}{2 \, a^2}  N_0 \pi_a {}\pi_{\lambda0} \lambda_0 = 0 
\end{eqnarray}

Equation (\ref{sc02}) leads to zeroth order Hamiltonian constraint and
is equivalent to equation (\ref{hc0l}). Further, conservation of
secondary constraint (\ref{sc01}) leads to tertiary constraint
\begin{eqnarray}
\label{tc0}
\Phi_t \equiv \{\Phi_s^1, \mathcal{H}_{D0}\} \approx 0 
&& \Rightarrow {}\pi_{\varphi0} - \kappa N_0^2 {}\pi_{\lambda0}{}^2 a - 3 \kappa N_0^2 {}\pi_{\lambda0}{}^3 \lambda_0 = 0
\end{eqnarray}
and generates quaternary constraint
\begin{eqnarray}
&&~~~~\Phi_q \equiv \{\Phi_t, \mathcal{H}_{D0}\} \approx 0 \\
&&\Rightarrow S_0 \left(- 2 \kappa N_0^2 {}\pi_{\lambda0} a + 15 \kappa N_0^5{}\pi_{\lambda0}{}^5 \pi_a a \right) - N_0 V_{\varphi} - \frac{5}{2} \kappa^2 N_0^3 {}\pi_{\lambda0}{}^3 \lambda_0 a^{-2}   \nonumber \\
&&~~~~ + \frac{5}{6} \kappa^2 N_0^3 {}\pi_{\lambda0}{}^2 \pi_a  a^{-1}+ 18 \kappa^2 N_0^6 {}\pi_{\lambda0}{}^6 \lambda_0  + 6 \kappa^2 N_0^6 {}\pi_{\lambda0}{}^6 \pi_a a = 0
\end{eqnarray}

Out of the 6 constrained equations, equations (\ref{pc02}) and
(\ref{sc02}) are first class and rest are second class
constraints. Hence, in coordinate space, number of degrees of freedom
is \[\frac{1}{2} \times (10 - 2 \times 2 - 4) = 1\] which is same as
canonical scalar field model.

\subsubsection{First order perturbed equations}
\label{galfirst}
The first order perturbation of the field variables and the momenta
are defined as
\begin{eqnarray}
\label{fog}
&&N = N_0 + \epsilon N_1, ~~~~ N^{i} = \epsilon N_1^{i},~~~~ \gamma_{i j} = a^2 \delta_{i j}, ~~~~S= S_0 + \epsilon S_1,\nonumber\\
&&\lambda=\lambda_0 + \epsilon \lambda_1, \varphi = \varphi_0 + \epsilon \varphi_1 ~~~~\pi^{i j} = {}\pi_0{}^{i j} + \epsilon \pi_1^{i j}, ~~~~ \pi_N = \pi_{N0} + \epsilon \pi_{N1},\nonumber\\
&&\pi_\varphi= {}\pi_{\varphi0} + \epsilon \pi_{\varphi1},~~~~\pi_\lambda = {}\pi_{\lambda0} + \epsilon \pi_{\lambda 1}.
\end{eqnarray}

First order perturbed Hamiltonian equations are obtained by varying
the second order perturbed Hamiltonian. Using above definitions of
perturbations in the Hamiltonian (\ref{DHamiltonian}), the second
order perturbed Hamiltonian becomes 
\begin{eqnarray}
\label{HG2}
&&\mathcal{H}_{D2} = - S_1 \lambda_1 + N_1 V_\varphi  a{}^{3} \varphi_1 + \frac{1}{2}\, N_0 V_{\varphi \varphi}  \varphi_1{}^{2} a{}^{3} + {N_1}^{i} {}\pi_{\lambda0} {\partial}_{i}{\lambda_1}\,  + {N_1}^{i} {}\pi_{\varphi0} {\partial}_{i}{\varphi_1}\,   -  \nonumber \\
&&2\, N_0 N_1 {}\pi_{\lambda0} \pi_{\varphi1} - 2\, N_0 N_1 \pi_{\lambda 1} {}\pi_{\varphi0} - \pi_{\lambda 1} \pi_{\varphi1} N_0{}^{2} + \pi_{\lambda 1} \lambda_0 {\partial}_{i}{{N_1}^{i}}\,  + {}\pi_{\lambda0} \lambda_1 {\partial}_{i}{{N_1}^{i}}\,   - \nonumber \\
&& {\delta}^{i j} {\partial}_{i}{\lambda_1}\,  {\partial}_{j}{\varphi_1}\,  a{}^{(-2)} - S_0 N_0{}^{3} \pi_{\lambda 1}{}^{2} a{}^{3} - 6\, N_1 {}\pi_{\lambda0} \pi_{\lambda 1} S_0 N_0{}^{2} a{}^{3} - 3\, N_0 S_0 N_1{}^{2} {}\pi_{\lambda0}{}^{2} a{}^{3} -\nonumber \\
&&2\, {}\pi_{\lambda0} \pi_{\lambda 1} S_1 N_0{}^{3} a{}^{3} - 3\, N_1 S_1 N_0{}^{2} {}\pi_{\lambda0}{}^{2} a{}^{3}%
 - \frac{3}{4}\, N_0 \kappa {}\pi_{\lambda0}{}^{2} \lambda_1{}^{2} a{}^{(-3)}  - \nonumber \\
&& \frac{3}{4}\, N_0 \kappa \pi_{\lambda 1}{}^{2} \lambda_0{}^{2} a{}^{(-3)} -\frac{3}{2}\, N_1 \kappa \lambda_0 \lambda_1 {}\pi_{\lambda0}{}^{2} a{}^{(-3)} - \frac{3}{2}\, N_1 {}\pi_{\lambda0} \pi_{\lambda 1} \kappa \lambda_0{}^{2} a{}^{(-3)}  + \nonumber \\
&&{\delta}^{i j} \lambda_0 {\partial}_{i}{N_1}\,  {\partial}_{j}{\varphi_1}\,  N_0{}^{-1} a{}^{(-2)} + N_0 S_0 {\delta}^{i j} {\partial}_{i}{\varphi_1}\,  {\partial}_{j}{\varphi_1}\,  a - N_0 \pi_{\lambda 1} {\delta}_{i j} \kappa \lambda_0 {\pi_1}^{i j} a{}^{-1} - \nonumber \\
&&N_1 {}\pi_{\lambda0} {\delta}_{i j} \kappa \lambda_0 {\pi_1}^{i j} a{}^{-1} - N_1 \pi_{\lambda 1} {\delta}_{i j} \kappa \lambda_0 \pi_0{}^{i j} a{}^{-1} - N_0 {}\pi_{\lambda0} {\delta}_{i j} \kappa \lambda_1 {\pi_1}^{i j} a{}^{-1} - \nonumber \\
&&N_1 {}\pi_{\lambda0} {\delta}_{i j} \kappa \lambda_1 \pi_0{}^{i j} a{}^{-1} - N_0 {\delta}_{i j} {\delta}_{k l} \kappa {\pi_1}^{i j} {\pi_1}^{k l} a - 2\, N_1 {\delta}_{i j} {\delta}_{k l} \kappa \pi_0{}^{i j} {\pi_1}^{k l} a  + \nonumber \\
&&4\, N_1 {\delta}_{i j} {\delta}_{k l} \kappa {{}\pi_0{}}^{i k} {\pi_1}^{j l} a %
 + \pi_{N1} \xi_1 - N_1 {}\pi_{\lambda0} \lambda_0 N_0{}^{(-2)} \xi_1 + {}\pi_{\lambda0} \lambda_0 N_0{}^{(-3)} N_1{}^{2} \xi_0  -\nonumber \\
&& N_1 \pi_{\lambda 1} \lambda_0 N_0{}^{(-2)} \xi_0 + {}\pi_{\lambda0} \lambda_1 N_0{}^{-1} \xi_1 -N_1 {}\pi_{\lambda0} \lambda_1 N_0{}^{(-2)} \xi_0 +\pi_{\lambda 1} \lambda_1 N_0{}^{-1} \xi_0\nonumber \\
&&- {}\pi_{\lambda0} {}\pi_{\varphi0} N_1{}^{2}+ 2 {\delta}_{i j} {\partial}_{k}{{N_1}^{i}}\,  {\pi_1}^{j k} a{}^{2}- 3\, N_0 {}\pi_{\lambda0} \pi_{\lambda 1} \kappa \lambda_0 \lambda_1 a{}^{(-3)}+ \pi_{\lambda 1} \lambda_0 N_0{}^{-1} \xi_1\nonumber \\
&&+ 2\, N_0 {\delta}_{i j} {\delta}_{k l} \kappa {\pi_1}^{i k} {\pi_1}^{j l} a- N_0 \pi_{\lambda 1} {\delta}_{i j} \kappa \lambda_1 \pi_0{}^{i j} a{}^{-1} - {N_1}^{i} {}\pi_{\lambda0} \lambda_0 {\partial}_{i}{N_1}\,  N_0{}^{-1}.
\end{eqnarray}

Equations of motion corresponding to the perturbed Hamiltonian (\ref{HG2}) is expressed in Appendix \ref{galfirst}.

Variation of the perturbed Hamiltonian density (\ref{HG2}) with
respect to $S_1$ lead to the secondary constraint
\begin{equation}
\label{lam1}
\lambda_1 =  - 2\, \pi_{\lambda 1} {}\pi_{\lambda0} N_0{}^{3} a{}^{3} - 3\, N_1 N_0{}^{2} {}\pi_{\lambda0}{}^{2} a{}^{3}.
\end{equation}

Variation of (\ref{HG2}) with respect to $\xi_1$ leads to the equation
\begin{equation}
\label{fcons1}
\pi_{N1} =  - \lambda_0 \pi_{\lambda 1} N_0{}^{-1} - \lambda_1 {}\pi_{\lambda0} N_0{}^{-1} + \lambda_0 {}\pi_{\lambda0} N_0{}^{(-2)} N_1
\end{equation}
which constrains $\pi_{N1}$ with $\pi_{\lambda 1}$. Since there is no
perturbation in the 3-metric, variation with respect to $\pi_1^{i j}$,
as expected, is equal to zero and contracting with $(\delta^{m n} \delta^{i j} -
\delta^{m i} \delta^{n j})$, we get
{\small
\begin{eqnarray}
\label{fcons2}
&& {\pi_1}^{i j} = \frac{1}{2}\, N_0{}^{-1} a \kappa{}^{-1} {\delta}^{i j} {\partial}_{k}{{N_1}^{k}}\,  - \frac{1}{2}\, a{}^{(-2)} {\delta}^{i j} \pi_{\lambda 1} \lambda_0 - \frac{1}{2}\, a{}^{(-2)} N_1 N_0{}^{-1} {\delta}^{i j} {}\pi_{\lambda0} \lambda_0 \nonumber \\
&&- \frac{1}{2}\, a{}^{(-2)} {\delta}^{i j} {}\pi_{\lambda0} \lambda_1 - \frac{1}{4}\, N_0{}^{-1} \kappa{}^{-1} a {\delta}^{k j} {\partial}_{k}{{N_1}^{i}}\,  - \frac{1}{4}\, N_0{}^{-1} \kappa{}^{-1} a {\delta}^{k i} {\partial}_{k}{{N_1}^{j}}\,  - N_0{}^{-1} N_1 \pi_0{}^{i j}~~~~~~~
\end{eqnarray}}

At first order perturbation, Hamilton's equations corresponding to the
time variation of field variables are given by {\small
\begin{eqnarray}
\label{del0lam1}
&& {\lambda_1{}^\prime}\,  = - 2\, N_0 N_1 {}\pi_{\varphi0} - \pi_{\varphi1} N_0{}^{2} + \lambda_0 {\partial}_{i}{{N_1}^{i}}\,  - 2\, \pi_{\lambda 1} S_0 N_0{}^{3} a{}^{3} - 6\, N_1 {}\pi_{\lambda0} S_0 N_0{}^{2} a{}^{3} - \nonumber \\
&&~~~~~~~~ 2\, {}\pi_{\lambda0} S_1 N_0{}^{3} a{}^{3} - 3\, N_0 {}\pi_{\lambda0} \kappa \lambda_0 \lambda_1 a{}^{(-3)} - \frac{3}{2}\, N_0 \pi_{\lambda 1} \kappa \lambda_0{}^{2} a{}^{(-3)} - \frac{3}{2}\, N_1 {}\pi_{\lambda0} \kappa \lambda_0{}^{2} a{}^{(-3)} \nonumber \\
&&~~~~~~~~ - N_0 {\delta}_{i j} \kappa \lambda_0 {\pi_1}^{i j} a{}^{-1} - N_1 {\delta}_{i j} \kappa \lambda_0 \pi_0{}^{i j} a{}^{-1} - N_0 {\delta}_{i j} \kappa \lambda_1 \pi_0{}^{i j} a{}^{-1} + \nonumber \\
&&~~~~~~~~\lambda_0 N_0{}^{-1} \xi_1 - N_1 \lambda_0 N_0{}^{(-2)} \xi_0 + \lambda_1 N_0{}^{-1} \xi_0 \\
&& \varphi_1^{\prime}\,  = - 2\, N_0 N_1 {}\pi_{\lambda0} - \pi_{\lambda 1} N_0{}^{2} \\
&& {N_1^\prime} = \xi_1
\end{eqnarray}}

Hamilton's equation corresponding to the time variation of the momentum
$\pi_{\lambda 1}$ is given by {\small
\begin{eqnarray}
&& {\pi_{\lambda 1}{}^\prime} + \frac{\delta  \mathcal{H}_{D2}}{\delta \lambda_1} = 0 \\
&& \Rightarrow {\pi_{\lambda 1}{}^\prime}\,  - S_1 + {\delta}^{i j} {\partial}_{i j}{\varphi_1}\,  a{}^{(-2)} - \frac{3}{2}\, N_0 \kappa {}\pi_{\lambda0}{}^{2} \lambda_1 a{}^{(-3)} - 3\, N_0 {}\pi_{\lambda0} \pi_{\lambda 1} \kappa \lambda_0 a{}^{(-3)} -  \nonumber \\
&&~~~~~~~~\, N_0 {}\pi_{\lambda0} {\delta}_{i j} \kappa {\pi_1}^{i j} a{}^{-1}- \frac{3}{2}\, N_1 \kappa \lambda_0 {}\pi_{\lambda0}{}^{2} a{}^{(-3)} -  N_0 \pi_{\lambda 1} {\delta}_{i j} \kappa \pi_0{}^{i j} a{}^{-1}   \nonumber \\
&&~~~~~~~~-  N_1 {}\pi_{\lambda0} {\delta}_{i j} \kappa \pi_0{}^{i j} a{}^{-1}  +{}\pi_{\lambda0} N_0{}^{-1} \xi_1 - N_1 {}\pi_{\lambda0} N_0{}^{(-2)} \xi_0 + \pi_{\lambda 1} N_0{}^{-1} \xi_0 = 0~~~~~~~~~~
\end{eqnarray}}
which, using other first order equations and (\ref{ncomp}), becomes
{\small
\begin{eqnarray}
&&- S_1 - \varphi_1^{\prime \prime}\,  a{}^{(-2)} + {\delta}^{i j} {\partial}_{i j}{\varphi_1}\,  a{}^{(-2)} + {\phi_1^\prime}\,  \varphi_0^{\prime}\,  a{}^{(-2)} - 2\, \varphi_1^{\prime}\,  a^{\prime}\,  a{}^{(-3)} + \nonumber \\
&&~~~~~~~~2\, \varphi_0^{\prime \prime}\,  \phi_1 a{}^{(-2)} + {\delta}^{i j} \varphi_0^{\prime}\,  {\partial}_{i j}{B_1}\,  a{}^{(-2)} + 4\, \varphi_0^{\prime}\,  a^{\prime}\,  \phi_1 a{}^{(-3)} = 0 \nonumber \\
&&\Rightarrow S_1 = - \varphi_1^{\prime \prime}\,  a{}^{(-2)} + {\delta}^{i j} {\partial}_{i j}{\varphi_1}\,  a{}^{(-2)} + {\phi_1^\prime}\,  \varphi_0^{\prime}\,  a{}^{(-2)} - 2\, \varphi_1^{\prime}\,  H\,  a{}^{(-2)}\nonumber \\
&&~~~~~~~~ + 2\, \varphi_0^{\prime \prime}\,  \phi_1 a{}^{(-2)} + {\delta}^{i j} {\partial}_{i j}{B_1}\,  \varphi_0^{\prime}\,  a{}^{(-2)} + 4\, \varphi_0^{\prime}\,  H\,  \phi_1 a{}^{(-2)}.
\end{eqnarray}}

Right hand side of the above equation is the explicit form of first
order perturbed $\Box \varphi$. Hence, the first equation obtained
from perturbed Hamiltonian is consistent.

First order perturbed Hamiltonian constraint is obtained by the time
variation of $\pi_{N1}$ and is given by {\small
\begin{eqnarray}
&&~~~~{\pi_{N1}{}^\prime} + \frac{\delta  \mathcal{H}_{D2}}{\delta N_1} = 0 \\
&& \Rightarrow {\pi_{N1}{}^\prime}\,  + V_\varphi  a{}^{3} \varphi_1 - 2\, N_1 {}\pi_{\lambda0} {}\pi_{\varphi0} - 2\, N_0 {}\pi_{\lambda0} \pi_{\varphi1} - 2\, N_0 \pi_{\lambda 1} {}\pi_{\varphi0} - 6\, {}\pi_{\lambda0} \pi_{\lambda 1} S_0 N_0{}^{2} a{}^{3} -   \nonumber \\
&&~~~~6\, N_0 N_1 S_0 {}\pi_{\lambda0}{}^{2} a{}^{3} -3\, S_1 N_0{}^{2} {}\pi_{\lambda0}{}^{2} a{}^{3} -\frac{3}{2}\, \kappa \lambda_0 \lambda_1 {}\pi_{\lambda0}{}^{2} a{}^{(-3)} - \frac{3}{2}\, {}\pi_{\lambda0} \pi_{\lambda 1} \kappa \lambda_0{}^{2} a{}^{(-3)}  \nonumber \\
&&~~~~+ {}\pi_{\lambda0} \lambda_0 {\partial}_{i}{{N_1}^{i}}\,  N_0{}^{-1} - {\delta}^{i j} \lambda_0 {\partial}_{i j}{\varphi_1}\,  N_0{}^{-1} a{}^{(-2)} - {}\pi_{\lambda0} {\delta}_{i j} \kappa \lambda_0 {\pi_1}^{i j} a{}^{-1} -  \nonumber \\
&&~~~~\pi_{\lambda 1} {\delta}_{i j} \kappa \lambda_0 \pi_0{}^{i j} a{}^{-1} - {}\pi_{\lambda0} {\delta}_{i j} \kappa \lambda_1 \pi_0{}^{i j} a{}^{-1} -2\, {\delta}_{i j} {\delta}_{k l} \kappa \pi_0{}^{i j} {\pi_1}^{k l} a + 4\, {\delta}_{i j} {\delta}_{k l} \kappa {{}\pi_0{}}^{i k} {\pi_1}^{j l} a  \nonumber \\
&&~~~~ - {}\pi_{\lambda0} \lambda_0 N_0{}^{(-2)} \xi_1 + 2\, N_1 {}\pi_{\lambda0} \lambda_0 N_0{}^{(-3)} \xi_0%
 -\pi_{\lambda 1} \lambda_0 N_0{}^{(-2)} \xi_0 -{}\pi_{\lambda0} \lambda_1 N_0{}^{(-2)} \xi_0 = 0~~~~~~~~~~
\end{eqnarray}}

In flat-slicing gauge, it becomes
\begin{eqnarray}
\label{hc1gh}
&&\mathcal{H}_{N1} \equiv 24\, H\,  \phi_1 \varphi_0^{\prime}\, {}^{3} a{}^{(-2)} - 18\, \varphi_1^{\prime}\,  H\,  \varphi_0^{\prime}\, {}^{2} a{}^{(-2)} + V_{\varphi}\,  a{}^{2} \varphi_1 + 2\, {\delta}^{i j} {\partial}_{i j}{B_1}\,  \varphi_0^{\prime}\, {}^{3} a{}^{-2} + \nonumber \\
&&2\, {\delta}^{i j} {\partial}_{i j}{\varphi_1}\,  \varphi_0^{\prime}\, {}^{2} a{}^{-2} + 2\, {\delta}^{i j} H\,  {\partial}_{i j}{B_1}\,  \kappa{}^{-1} + 6\, \phi_1 \kappa{}^{-1} H^2 = 0.
\end{eqnarray}

Since there is no momentum $\pi^1_i$ corresponding to $N_1^{i}$
appeared in the second order Hamiltonian (\ref{HG2}), variation of the
Hamiltonian with respect to $N_1^{i}$ leads to Momentum constraint of
Galilean field
\begin{equation}
\label{mcg1}
M_{i1} \equiv {}\pi_{\varphi0} {\partial}_{i}{\varphi_1}\,  - \lambda_0 {\partial}_{i}{\pi_{\lambda 1}}\,  - 2\, {\delta}_{i j} a{}^{2} {\partial}_{k}{{\pi_1}^{j k}}\,  - {}\pi_{\lambda0} \lambda_0 {\partial}_{i}{N_1}\,  N_0{}^{-1} = 0,
\end{equation}
which, in flat-slicing gauge, becomes
\begin{equation}
\label{mc1gh}
M_{i1} \equiv - 6\, H\,  {\partial}_{i}{\varphi_1}\,  \varphi_0^{\prime}\, {}^{2}  - 2\, {\partial}_{i}{\phi_1}\,  \varphi_0^{\prime}\, {}^{3} + 2\, {\partial}_{i}{\varphi_1{}^\prime}\,  \varphi_0^{\prime}\, {}^{2} - 2\, H\,  {\partial}_{i}{\phi_1}\,  \kappa{}^{-1} a^2 = 0.
\end{equation}

Similarly, equation of motion of $\varphi_1$ can be obtained from
\begin{eqnarray}
&&~~~~{\pi_{\varphi1}{}^\prime} + \frac{\delta \mathcal{H}_{D2}}{\delta \varphi_1} = 0 \\
&& \Rightarrow {\pi_{\varphi1}{}^\prime}\,  + N_1 V_{\varphi}\,  a{}^{3} + N_0 V_{\varphi \varphi}\,  \varphi_1 a{}^{3} - {\partial}_{i}{{N_1}^{i}}\,  {}\pi_{\varphi0} + {\delta}^{i j} {\partial}_{i j}{\lambda_1}\,  a{}^{(-2)} \nonumber \\
&& ~~~~~~~~- {\delta}^{i j} \lambda_0 {\partial}_{i j}{N_1}\,  N_0{}^{-1} a{}^{(-2)} -2\, N_0 S_0 {\delta}^{i j} {\partial}_{i j}{\varphi_1}\,  a  = 0.
\end{eqnarray}

Using (\ref{ncomp}), the above equation can be written as
\begin{eqnarray}
\label{eomphi1gh}
&& - 18\, \phi_1 \varphi_0^{\prime}\, {}^{2} H^2 + 12\, \varphi_0^{\prime}\,  \varphi_1^{\prime}\,  H^2 + 18\, \phi_1^\prime\,  H\,  \varphi_0^{\prime}\, {}^{2}  \nonumber \\
&&+ 36\, \varphi_0^{\prime}\,  H\,  \varphi_0^{\prime \prime}\,  \phi_1  - 12\, \varphi_0^{\prime}\,  H\,  \varphi_1^{\prime \prime} - 12\, \varphi_1^{\prime}\,  H\,  \varphi_0^{\prime \prime} \nonumber \\
&& - 12\, \varphi_0^{\prime}\,  \varphi_1^{\prime}\,  a^{\prime \prime}\,  a{}^{-1} + 6\, {\delta}^{i j} H\,  {\partial}_{i j}{B_1}\,  \varphi_0^{\prime}\, {}^{2}  + 4\, {\delta}^{i j} \varphi_0^{\prime}\,  \varphi_0^{\prime \prime}\,  {\partial}_{i j}{B_1}\,  + \nonumber \\
&&4\, {\delta}^{i j} \varphi_0^{\prime}\,  H\,  {\partial}_{i j}{\varphi_1}\,  a{}^{-1} + 4\, {\delta}^{i j} \varphi_0^{\prime \prime}\,  {\partial}_{i j}{\varphi_1}\,  + 2\, {\delta}^{i j} {\partial}_{i j}{B_1^\prime}\,  \varphi_0^{\prime}\, {}^{2} + V_{\varphi}\,  \phi_1 a{}^{4} + \nonumber \\
&&V_{\varphi \varphi}\,  a{}^{4} \varphi_1 + 2\, {\delta}^{i j} {\partial}_{i j}{\phi_1}\,  \varphi_0^{\prime}\, {}^{2}+ 18\, a^{\prime \prime}\,  \phi_1 \varphi_0^{\prime}\, {}^{2} a{}^{-1} = 0.
\end{eqnarray}

Equations (\ref{hc1gh}), (\ref{mc1gh}) and (\ref{eomphi1gh}) obtained
from Hamiltonian formulation are identical to (\ref{hc1ga}),
(\ref{mc1ga}) and (\ref{eomphi1ga}), respectively. 

Third and fourth order interaction Hamiltonian are given in Appendix \ref{intHamGal}.

\paragraph{\underline{Counting scalar degrees of freedom at first order:}}
Since we are considering only scalar first order perturbations,
$N_1^i$ contains one scalar variable ($N_1^i = \delta^{i
  j} \partial_{j}{B_1}$) and hence, $\pi_i = 0$ and Momentum
constraint (\ref{mcg1}) lead to two constrained equations. Along with
(\ref{fcons1}) and Hamiltonian constraint (\ref{hc1gh}), we get 4
constrained equations. Similarly, as we have seen in zeroth order, at
first order, we get four second class constraints:
\begin{eqnarray}
&& \Phi_p \equiv \pi_S = 0 \\
&& \Phi_s \equiv \{\pi_S, \mathcal{H}_{D2}\} = \lambda_1 + 2\, \pi_{\lambda 1} {}\pi_{\lambda0} N_0{}^{3} a{}^{3} + 3\, N_1 N_0{}^{2} {}\pi_{\lambda0}{}^{2} a{}^{3} = 0\\
\label{fcons3}
&&\Phi_t \equiv \{\Phi_s, \mathcal{H}_{D2}\} = \pi_{\varphi1} + 2\, N_1 {}\pi_{\varphi0} N_0{}^{-1} - 2\, N_0 {\partial}_{i}{{N_1}^{i}}\,  {}\pi_{\lambda0}{}^{2} a{}^{3} + 2\, N_0 {}\pi_{\lambda0} {\delta}^{i j} {\partial}_{i j}{\varphi_1}\,  a + \nonumber \\
&&~~~~~~~~9\, N_1 \kappa N_0{}^{4} {}\pi_{\lambda0}{}^{5} a{}^{3} +  6\, \pi_{\lambda 1} \kappa N_0{}^{5} {}\pi_{\lambda0}{}^{4} a{}^{3} -3\, N_0 N_1 \pi_a \kappa {}\pi_{\lambda0}{}^{2} a \nonumber \\
&&~~~~~~~~- 2\, \pi_a {}\pi_{\lambda0} \pi_{\lambda 1} \kappa N_0{}^{2} a = 0~~~\\
&& \Phi_q \equiv \{\Phi_t, \mathcal{H}_{D2}\} \approx 0
\end{eqnarray}

We also get 12 more second class constraints: $\delta\gamma_{i j} = 0$
and $\{\delta\gamma_{i j}, ^{(2)}\mathcal{H}_D\}$ which leads to
equation (\ref{fcons2}).  Since, we have fixed the gauge, all
constraints become second class. Our Galilean phase space contains 22
variables. Hence in configuration space, the number of degrees of
freedom is \[\frac{1}{2}\times(22 - 4 - 4- 12)= 1.\]

This procedure can be extended to higher order and at 
any order it can be shown that the degrees of freedom is one. 
So, at any order, Galilean scalar field
produces no extra degrees of freedom due to the higher derivative
terms present in the Lagrangian and behave exactly same as any single
derivative Lagrangian system.

However, if we consider generalized Lagrangian containing second order 
derivative terms, the above analysis can not be extended. In that 
case, unlike Galilean field, the Lapse function and Shift vector 
will not act like constraints and hence, 
$\pi_N + ... \neq 0,~\pi_i + ...\neq 0$ since
$\pi_N$ and $\pi_i$ contain time derivatives of $N$ and
$N^{i}$\cite{Nandi:2015tha}, and dynamical degrees of freedom can be
generated from those. This means that for any higher derivative
gravitational theory, lack of first class constraints lead to extra
degrees of freedom. Similarly, extra degrees of freedom will always be
generated for any generalized second order derivative Lagrangian.

\section{Conclusion and Discussion}

In this work, using Hamiltonian formulation, we have formulated 
a consistent cosmological perturbation theory at all orders. We
have adopted the following procedures: we choose a particular gauge
that does not lead to any particular gauge artifact\cite{Malik2009}
such that some variables remain unperturbed while others can be
separated as zeroth order part and perturbation part. In order to make
the procedure transparent, we considered a simple model of two
variables where one variable is unperturbed and other variable can be
perturbed. At first order, we confronted the gauge-issue and found
that, even canonical conjugate momentum of unperturbed quantity has
perturbation part that leads a constrained equation at every perturbed
order and by using the equation we can get the exact form of perturbed
momentum. We fixed the gauge-issue and obtained all first order
perturbed equations as well as third and fourth order perturbed
Hamiltonian which is consistent with Lagrangian formulation. The
procedure is simple and robust and can be extended to any order of
perturbation. Table below provides a bird's eye view of the both the formulations and 
advantages of the Hamiltonian formulation that is proposed in this work:

\begin{center}
\begin{tabular}{|m{2cm}|m{6cm}|m{6cm}|}
\hline
 & Lagrangian formulation & Hamiltonian formulation \\
\hline
Gauge conditions and {guage-invariant equations} & At any order, choose a gauge {which} does not lead to gauge-artifacts & 
Choose a gauge {with no gauge-artifacts}, however, momentum corresponding  to unperturbed quantity is non-zero leading to consistent equations of motion.\\
\hline
Dynamical variables & {Counting true dynamical degrees of freedom is difficult.} & Using Dirac's procedure, constraints can 
easily be obtained and is easy to determine the degrees of freedom.\\
\hline
Quantization at all orders &  Difficult to quantize constrained system. & Since constraints are obtained systematically and reduced phase space contains only true degrees of freedom, it is straightforward to quantize the theory using Hamiltonian formulation.\\
\hline
Calculating the observables & Requires to invert the expressions at each order and hence non-trivial 
to compute higher-order correlation function. & Once the relation between $\varphi$ and Curvature perturbation\footnotemark[4]  is known, 
calculating the correlation functions from the Hamiltonian is simple and straightforward to obtain.\\
\hline
\end{tabular}
\end{center}

\footnotetext[4]{It is important to note that, in the case of first order, relation between $\varphi$ and three-curvature is straight forward. However, it is more subtle in the case of higher-order perturbations\cite{Malik2009}.}
We have applied the procedure to canonical scalar field minimally
coupled to gravity and similarly obtained all equations as well as
interaction Hamiltonian using both Lagrangian and Hamiltonian
formulation. Both lead to identical results. We also showed that,
obtaining interaction Hamiltonian by using Hamiltonian formulation is efficient and straightforward. Unlike the Lagrangian 
formulation, we do not need to invert the expressions at each order~\cite{Huang:2006eha}.

We, then obtained a consistent perturbed Hamiltonian formulation for
Galilean scalar fields. Using flat-slicing gauge, we have obtained zeroth and 
first order Hamilton's equations that are consistent with Lagrangian formulation. 
We carefully analyzed the constraints in the system and counted degrees of freedom at 
every order in the system that is consistent with the results of Deffayet \textit{et
  al}\cite{Deffayet:2015qwa} results. It has been shown that general higher derivative
models lead to dynamical equations of Lapse function/shift vector which
increases the number degrees of freedom of the system. But, only in
higher derivative Galilean theory, Lapse function/shift vector remain
constraint, therefore, no extra degrees of freedom flows in the
system. Similar type of problem has been encountered in a different manner in Ref. \cite{Nandi:2015tha}.

To make the Physics transparent, in this work, we have  neglected vector and tensor perturbations.  Our 
approach can be applied to higher-order perturbations including vector and tensor perturbations. In the 
presence of tensor or vector or any mixed modes at any higher perturbed order, since modes do not 
decouple, $\pi^{i j}$ cannot decouple and act as the momentum corresponding the overall 3-metric 
$\delta \gamma_{i j}$ which contains mixed modes. Hence, 
\begin{eqnarray}
&& \frac{\delta \mathcal{H}}{\delta \pi^{i j} } = \partial_0 (\delta \gamma_{i j}) \nonumber \\
&& ~~~~~~= \partial_0 (\delta \gamma^S_{i j} + \delta \gamma^V_{i j} + \delta \gamma^T_{i j}), \nonumber 
\end{eqnarray}
where $\gamma^S_{i j}, \gamma^V_{i j}, \gamma^T_{i j}$ are scalar, vector and tensor modes, respectively.

Our approach can be applied to any model of gravity and matter fields to obtain any higher order interaction Hamiltonian without invoking any approximation such as slow roll, etc. It can also been shown that, the mechanism can be applied for any generalized tensor fields and it can even extract any higher order cross-correlation interaction Hamiltonian.

{Our approach can also be used for modified gravity models including $f(R)$ model, other scalar-tensor theories like Gauss-Bonnet inflation, Lovelock gravity,
Hordenski theory. In fact, we can say unequivocally that our approach can be used for any kind of gravity models.
}

\section{Acknowledgements}
We thank the support of Max Planck-India Partner group on Gravity and
Cosmology.  DN is supported by CSIR fellowship. SS is partially
supported by Ramanujan Fellowship of DST, India. Further we thank
Kasper Peeters for his useful program
Cadabra\cite{Peeters:2007wn,DBLP:journals/corr/abs-cs-0608005} and
useful algebraic calculations with it.

\appendix
\section{Perturbed equations of motion of Canonical scalar field in flat-slicing gauge}\label{PLeCSF}
 
\subsection{Background equations}

0-0 component of the Einstein's equation or the Hamiltonian constraint
in conformal coordinate is given by
\begin{equation}
\label{B00EE}
 {H}^2 \equiv a^{\prime}{}^{2} a^{-2} = \frac{\kappa}{3} \left[\frac{1}{2}\,  \varphi_0^{\prime}\, {}^{2} + V a^2 \right].
\end{equation}

Trace of i-j component of the Einstein's equation gives the equation
of motion of $a$, which can also be obtained by varying the zeroth
order action with respect to $a$ and is given by
\begin{equation}
\label{BijEE}
3\, \kappa{}^{-1} H^2 - 6\, \frac{a^{\prime \prime}}{a}\,  \kappa{}^{-1} - \frac{3}{2}\, \varphi_0^{\prime}\, {}^{2}  + 3\, V a{}^{2} = 0
\end{equation}
and the equation of motion of scalar field at zeroth order takes the
form
\begin{equation}
\label{BEoMS}
\varphi_0^{\prime \prime} + 2 {H}\, \varphi_0^{\prime}+ V_\varphi\,  a{}^{2} = 0.
\end{equation}

\subsection{First order perturbed equations}
Using the perturbed metric and perturbed scalar field defined in
(\ref{00pmetric}), (\ref{0ipmetric}), (\ref{3metric}) and
(\ref{field}), perturbed Hamiltonian constraint or the first order
perturbed 0-0 component of the Einstein's equation becomes
\begin{equation}
\label{100EE}
2\, {\delta}^{i j} H\,  {\partial}_{i j}{B_1}\,  \kappa{}^{-1} + 6\, \phi_1 \kappa{}^{-1} H^2 + \varphi_0^{\prime}\,  \varphi_1^{\prime} - \phi_1 \varphi_0^{\prime}\, {}^{2}  + V_\varphi \,  a{}^{2} \varphi_1 = 0.
\end{equation}

Similarly, perturbed Momentum constraint or perturbed 0-i component of
the Einstein's equation is
\begin{eqnarray}
\label{10iEE}
&& ~~\frac{2 H}{\kappa} \partial_{i}{\phi_1} = \varphi_0^\prime ~ \partial_{i}{\varphi_1} \\
&& \Rightarrow \phi_1 = \frac{\kappa}{2 {H}} \varphi_0^\prime ~\varphi_1
\end{eqnarray}
and equation of motion of the scalar field $\varphi_1$, using
(\ref{BEoMS}) is given by
\begin{equation}
\label{1EoMS}
\varphi_1^{\prime \prime}\, + 2\,H \varphi_1^\prime  - \phi_1^\prime\,  \varphi_0^{\prime}\,  + 2\, V_{\varphi}\,  \phi_1 a{}^{2} - {\delta}^{i j} \varphi_0^{\prime}\,  {\partial}_{i j}{B_1}\,  - {\delta}^{i j} {\partial}_{i j}{\varphi_1}\,  + V_{\varphi \varphi}\,  a{}^{2} \varphi_1 = 0.
\end{equation}

\subsection{Third order interaction Hamiltonian of Canonical scalar field  in terms of phase-space variables in flat-slicing gauge}\label{ThirdOrderInteractionCanonical}

Third or higher order perturbed Hamiltonian in terms of first order
perturbed variables is needed to calculate the interaction Hamiltonian
which helps to calculate higher order correlation functions. The third
order Hamiltonian is obtained by expanding the Hamiltonian (\ref{HamiltonianCanonical}) up to third order and is given by

\begin{eqnarray}
&& \mathcal{H}_3^C\, (\pi, \varphi) = - \frac{1}{2}\, N_1 {\partial}_{i}{{N_1}^{i}}\,  {\partial}_{j}{{N_1}^{j}}\,  N_0{}^{(-2)} \kappa{}^{-1} a{}^{3} + 2\, {\delta}_{i j} {\partial}_{k}{{N_1}^{k}}\,  \pi_0{}^{i j} N_0{}^{(-2)} N_1{}^{2} a{}^{2} - \nonumber \\
&&~~~~~~~~~~2\, {\delta}_{i j} {\delta}_{k l} \kappa \pi_0{}^{i j} {{}\pi_0{}}^{k l} N_0{}^{(-2)} N_1{}^{3} a + \frac{1}{4}\, N_1 {\delta}_{i j} {\delta}^{l k} {\partial}_{k}{{N_1}^{i}}\,  {\partial}_{l}{{N_1}^{j}}\,  N_0{}^{(-2)} \kappa{}^{-1} a{}^{3} + \nonumber \\
&&~~~~~~~~~~\frac{1}{4}\, N_1 {\partial}_{i}{{N_1}^{j}}\,  {\partial}_{j}{{N_1}^{i}}\,  N_0{}^{(-2)} \kappa{}^{-1} a{}^{3} + \frac{1}{2}\, {\delta}_{i j} {\partial}_{k}{{N_1}^{i}}\,  {{}\pi_0{}}^{k j} N_0{}^{(-2)} N_1{}^{2} a{}^{2} + \nonumber \\
&&~~~~~~~~~~\frac{1}{2}\, {\delta}_{i j} {\partial}_{k}{{N_1}^{i}}\,  {{}\pi_0{}}^{j k} N_0{}^{(-2)} N_1{}^{2} a{}^{2} - 2\, {\delta}_{i j} {\partial}_{k}{{N_1}^{k}}\,  \pi_0{}^{i j} N_0{}^{(-2)} N_1{}^{2} a{}^{2} + \nonumber \\
&&~~~~~~~~~~\frac{1}{2}\, {\delta}_{i j} {\partial}_{k}{{N_1}^{j}}\,  {{}\pi_0{}}^{k i} N_0{}^{(-2)} N_1{}^{2} a{}^{2} + \frac{1}{2}\, {\delta}_{i j} {\partial}_{k}{{N_1}^{j}}\,  {{}\pi_0{}}^{i k} N_0{}^{(-2)} N_1{}^{2} a{}^{2} + \nonumber \\
&&~~~~~~~~~~  2\, {\delta}_{i j} {\delta}_{k l} \kappa {{}\pi_0{}}^{i k} {{}\pi_0{}}^{j l} N_0{}^{(-2)} N_1{}^{3} a + \frac{1}{2}\, N_1 \pi_{\varphi1}{}^{2} a{}^{(-3)} + {N_1}^{i} \pi_{\varphi1} {\partial}_{i}{\varphi_1}\,  + \nonumber \\
&&~~~~~~~~~~{\delta}_{i j} {\delta}_{k l} \kappa \pi_0{}^{i j} {{}\pi_0{}}^{l k} N_0{}^{(-2)} N_1{}^{3} a + \frac{1}{2}\, N_1 {\delta}^{i j} {\partial}_{i}{\varphi_1}\,  {\partial}_{j}{\varphi_1}\,  a + \frac{1}{2}\, N_1 V_{\varphi \varphi}\,  \varphi_1{}^{2} a{}^{3} + \nonumber \\
&&~~~~~~~~~~\frac{1}{6}\, N_0 V_{\varphi \varphi \varphi}\,  \varphi_1{}^{3} a{}^{3}
\end{eqnarray}

\section{Hamiltonian formulation of Canonical scalar field in uniform density gauge}\label{udgcsf}
In uniform-density gauge, $E = \delta\varphi = 0$ and $\gamma_{i j} = a^2 (1 - 2 \epsilon {}\psi_1) \delta_{i j}, \gamma^{i j} = a^2 (1 + 2 \epsilon {}\psi_1 + 4 \epsilon^2 {}\psi_1^2) \delta^{i j}, \sqrt{\gamma} = a^6 ( 1 - 6 \epsilon {}\psi_1 + 12 \epsilon^2 {}\psi_1^2)$. The second order perturbed Hamiltonian is obtained by expanding the  Hamiltonian (\ref{HamiltonianCanonical}) up to second order by using above definitions and is given by

\begin{eqnarray}
\label{hamuniden}
&& \mathcal{H}_2^C = {\delta}_{i j} {\partial}_{k}{{N_1}^{j}}\,  {\pi_1}^{i k} a{}^{2} - 2\, {\delta}_{i j} {\partial}_{k}{{N_1}^{j}}\,  {{}\pi_0{}}^{i k} a{}^{2} {}\psi_1  -  2\, {N_1}^{i} {\delta}_{j k} {\partial}_{i}{{}\psi_1}\,  {{}\pi_0{}}^{j k} a{}^{2} +  \nonumber \\
&&{\delta}_{i j} {\partial}_{k}{{N_1}^{j}}\,  {\pi_1}^{i k} a{}^{2} -2\, {\delta}_{i j} {\partial}_{k}{{N_1}^{j}}\,  {{}\pi_0{}}^{i k} a{}^{2} {}\psi_1 - N_0 {\delta}_{i j} {\delta}_{k l} \kappa {\pi_1}^{i j} {\pi_1}^{k l} a - \nonumber \\
&&2\, N_1 {\delta}_{i j} {\delta}_{k l} \kappa \pi_0{}^{i j} {\pi_1}^{k l} a + 2\, N_0 {\delta}_{i j} {\delta}_{k l} \kappa \pi_0{}^{i j} {\pi_1}^{k l} {}\psi_1 a + N_1 {\delta}_{i j} {\delta}_{k l} \kappa \pi_0{}^{i j} {{}\pi_0{}}^{k l} {}\psi_1 a + \nonumber \\
&& \frac{1}{2}\, N_0 {\delta}_{i j} {\delta}_{k l} \kappa \pi_0{}^{i j} {{}\pi_0{}}^{k l} {}\psi_1{}^{2} a + 2\, N_0 {\delta}_{i j} {\delta}_{k l} \kappa {\pi_1}^{i k} {\pi_1}^{j l} a + 4\, N_1 {\delta}_{i j} {\delta}_{k l} \kappa {{}\pi_0{}}^{i k} {\pi_1}^{j l} a   + \nonumber \\
&&N_1 {}\pi_{\varphi0} \pi_{\varphi1} a{}^{(-3)} + 3\, N_0 {}\pi_{\varphi0} \pi_{\varphi1} a{}^{(-3)} {}\psi_1%
 + \frac{3}{2}\, N_1 {}\pi_{\varphi0}{}^{2} a{}^{(-3)} {}\psi_1 + \frac{15}{4}\, N_0 {}\pi_{\varphi0}{}^{2} {}\psi_1{}^{2} a{}^{(-3)}  - \nonumber \\
&&2\, N_0 {\delta}^{i j} {\partial}_{i j}{{}\psi_1}\,  \kappa{}^{-1} {}\psi_1 a - 2\, N_1 {\delta}^{i j} {\partial}_{i j}{{}\psi_1}\,  \kappa{}^{-1} a - 3\, N_1 V_0 a{}^{3} {}\psi_1 + \frac{3}{2}\, N_0 V_0 {}\psi_1{}^{2} a{}^{3}\nonumber \\
&&- 2\, N_1 {\delta}_{i j} {\delta}_{k l} \kappa {{}\pi_0{}}^{i k} {{}\pi_0{}}^{j l} {}\psi_1 a - N_0 {\delta}_{i j} {\delta}_{k l} \kappa {{}\pi_0{}}^{i k} {{}\pi_0{}}^{j l} {}\psi_1{}^{2} a + \frac{1}{2}\, N_0 \pi_{\varphi1}{}^{2} a{}^{(-3)}\nonumber \\
&&-4\, N_0 {\delta}_{i j} {\delta}_{k l} \kappa {{}\pi_0{}}^{i k} {\pi_1}^{j l} {}\psi_1 a- 3\, N_0 {\delta}^{i j} {\partial}_{i}{{}\psi_1}\,  {\partial}_{j}{{}\psi_1}\,  \kappa{}^{-1} a.
\end{eqnarray}

Since, $\varphi$ is unperturbed, variation of (\ref{hamuniden}) with respect to $\pi_{\varphi1}$ vanishes. 
\begin{eqnarray}
&&~~~\frac{\delta \mathcal{H}_2^C}{\delta \pi_{\varphi1}} = 0 \nonumber \\
&&\Rightarrow \pi_{\varphi1} = - \frac{N_1}{N_0} {}\pi_{\varphi0} - 3 {}\pi_{\varphi0} {}\psi_1
\end{eqnarray}

Explicit expression of $\pi_1^{i j}$ is obtained by varying the above Hamiltonian with respect to $\pi_1^{i j}$.
\begin{eqnarray}
&&~~~\partial_{0}\gamma_{i j} = \frac{\delta \mathcal{H}_2^C}{\delta \pi_1^{i j}} \nonumber \\
&&\Rightarrow \pi_1^{i j} = \kappa{}^{-1} (\frac{1}{2}\, N_0{}^{-1} a {\delta}^{i j} {\partial}_{k}{{N_1}^{k}}\,  + 2\, N_0{}^{-1} {\delta}^{i j} {\partial}_{0}{a}\,  {}\psi_1 + N_0{}^{-1} a {\delta}^{i j} {\partial}_{0}{\psi_1}\,  - \frac{1}{2}\, N_0{}^{-1} a {\delta}^{k j} {\partial}_{k}{{N_1}^{i}}\, )\nonumber \\
&&~~~~~~~~~~~ - N_0{}^{-1} N_1 \pi_0{}^{i j} + \pi_0{}^{i j} {}\psi_1
\end{eqnarray}

Using the above definitions and varying the Hamiltonian (\ref{hamuniden}) with respect to $N_1$, we get the Hamiltonian Constraint, which, in conformal coordinate becomes

\begin{eqnarray}
&&~~~\frac{\delta \mathcal{H}_2^C}{\delta N_1} = 0 \nonumber \\
&&\Rightarrow - 2\, {\delta}_{i j} {\delta}_{k l} \kappa \pi_0{}^{i j} {\pi_1}^{k l} a + {\delta}_{i j} {\delta}_{k l} \kappa \pi_0{}^{i j} {{}\pi_0{}}^{k l} {}\psi_1 a + 4\, {\delta}_{i j} {\delta}_{k l} \kappa {{}\pi_0{}}^{i k} {\pi_1}^{j l} a - 2\, {\delta}_{i j} {\delta}_{k l} \kappa {{}\pi_0{}}^{i k} {{}\pi_0{}}^{j l} {}\psi_1 a + \nonumber \\
&&~~~~~~~~~~{}\pi_{\varphi0} \pi_{\varphi1} a{}^{(-3)} + \frac{3}{2}\, {}\pi_{\varphi0}{}^{2} a{}^{(-3)} {}\psi_1 - 2\, {\delta}^{i j} {\partial}_{i j}{{}\psi_1}\,  \kappa{}^{-1} a - 3\, V_0 a{}^{3} {}\psi_1 = 0 \nonumber \\
&&\Rightarrow - 3 H \psi_1^\prime +  \delta^{i j} \partial_{i j}{}\psi_1 -  H \delta^{i j} \partial_{i j}B_1 -  \kappa \phi_1 V_0 a^2 = 0.
\end{eqnarray}

Similarly, varying the Hamiltonian with respect to $N_1^i$ leads to Momentum constraint and in conformal coordinate, it becomes

\begin{eqnarray}
&&~~~ \frac{\delta \mathcal{H}_2^C}{\delta N_1^{i}} = 0 \nonumber \\
&&\Rightarrow - 2 \delta_{i j} \partial_i \pi_1^{j k} + 4 \delta_{i j} \partial_k \psi_1 {}\pi_0{}^{i j} - 2 \delta_{j k} {}\pi_0{}^{j k} \partial_i \psi_1 = 0 \nonumber \\
&& \Rightarrow \partial_{i}{\psi_1^\prime} + H \partial_i \phi_1 = 0.
\end{eqnarray}

Finally, the equation of motion of $\psi_1$ is given by
\begin{eqnarray}
&&~~~\delta^{i j} \left(\partial_{0}{\pi_1^{i j}} + \frac{\delta \mathcal{H}_2^C}{\delta \gamma_{i j}}\right) = 0 \nonumber \\
&&\Rightarrow \delta^{i j}\left(\partial_{0}{\pi_1^{i j}} + \frac{\delta \mathcal{H}_2^C}{\delta \psi_1} \frac{\partial \psi}{\partial \gamma_{i j}}\right) = 0 \nonumber \\
&& \Rightarrow 2\, {\delta}^{i j} H\,  {\partial}_{i j}{B_1} + {\delta}^{i j} {\partial}_{i j}{B_1^\prime}\,   + 6H\, {\psi^\prime} + 3\, \psi^{\prime \prime}\,   + 3\kappa
\, V_0 \phi_1 a{}^{2} +  \nonumber \\
&&~~~~3H\, {\phi_1^\prime} - {\delta}^{i j} {\partial}_{i j}{\psi}\,   + {\delta}^{i j} {\partial}_{i j}{\phi_1}\,   = 0.
\end{eqnarray}

It can be verified that above equations are consistent with Lagrangian equations of motion.

\section{Perturbed Lagrangian equations of motion of Galilean scalar field in flat-slicing gauge}\label{galmod}

\subsection{Background equations}
In flat-slicing gauge, using (\ref{perturbfm1}) and
(\ref{perturbfm2}), (\ref{3metric}) and (\ref{field}), we get the
zeroth order Lagrangian density of the Galilean field (\ref{action})
\begin{equation}
\label{lb}
 \mathcal{L}_{G0} = - 3\, N_0{}^{-1} \kappa{}^{-1} a^{\prime}\, {}^{2} a - 2 a^{\prime}\,  N_0{}^{(-3)} \varphi_0^{\prime}\, {}^{3} a{}^{2} - N_0 V_0 a{}^{3}.
\end{equation}

Varying the Lagrangian (\ref{lb}) with respect to the
zeroth order Lapse function $N_0$, we get Hamiltonian constraint at
zeroth order and it is given by
\begin{equation}
\label{hc0l}
 3 N_0^{-2} \kappa^{-1} {a^\prime}{}^2 a + 6 N_0^{-4} \varphi_0^\prime \,{}^3 a^2 a^\prime - V_0 a^3 = 0.
\end{equation}

In conformal coordinate, the above equation takes the form
\begin{equation}
\label{hc0g1}
V_0 a{}^{2} - 6\, H a{}^{-2} \varphi_0^{\prime}\, {}^{3} - 3\,  \kappa{}^{-1} H^{2} = 0.
\end{equation}

Similarly, equation of motion of $a$ in conformal coordinate is given by
\begin{equation}
\label{eomag1}
6\, H\,  \varphi_0^{\prime}\, {}^{3} a{}^{(-2)} - 6\, \varphi_0^{\prime \prime}\,  \varphi_0^{\prime}\, {}^{2} a{}^{-2} + 3\, \kappa{}^{-1} H^2 - 6\, \frac{a^{\prime \prime}}{a}\,  \kappa{}^{-1} + 3\, V_0 a{}^{2}= 0
\end{equation}
and equation of motion of $\varphi_0$ in conformal coordinate is given by
\begin{equation}
\label{eomphig1}
 \frac{a^{\prime \prime}}{2\, a\, H}\,\varphi_0^{\prime}{}^2\,   - \frac{1}{2}\,H \varphi_0^{\prime}{}^2 + \varphi_0^{\prime \prime} \varphi_0^\prime\,  - \frac{1}{12\, H}\, V_{\varphi}\,   a{}^{4} = 0.
\end{equation}

\subsection{First order perturbation}\label{firstGaleq}

In conformal coordinate, it can be shown that, the equation of motion
of $N_1$ or the first order Hamiltonian constraint is
\begin{eqnarray}
\label{hc1ga}
&&24\, H\,  \phi_1 \varphi_0^{\prime}\, {}^{3} a{}^{(-2)} - 18\, \varphi_1^{\prime}\,  H\,  \varphi_0^{\prime}\, {}^{2} a{}^{(-2)} + V_{\varphi}\,  a{}^{2} \varphi_1 + 2\, {\delta}^{i j} {\partial}_{i j}{B_1}\,  \varphi_0^{\prime}\, {}^{3} a{}^{-2} + \nonumber \\
&&2\, {\delta}^{i j} {\partial}_{i j}{\varphi_1}\,  \varphi_0^{\prime}\, {}^{2} a{}^{-2} + 2\, {\delta}^{i j} H\,  {\partial}_{i j}{B_1}\,  \kappa{}^{-1} + 6\, \phi_1 \kappa{}^{-1} H = 0.
\end{eqnarray}

Similarly, equation of motion of $N_1^{i}$ or the first order
perturbed Momentum constraint is the following
\begin{equation}
\label{mc1ga}
- 6\, H\,  {\partial}_{i}{\varphi_1}\,  \varphi_0^{\prime}\, {}^{2}  - 2\, {\partial}_{i}{\phi_1}\,  \varphi_0^{\prime}\, {}^{3} + 2\, {\partial}_{i}{\varphi_1^\prime}\,  \varphi_0^{\prime}\, {}^{2} - 2\, H\,  {\partial}_{i}{\phi_1}\,  \kappa{}^{-1} a^2 = 0
\end{equation}
and equation of motion of $\varphi_1$ is given by
\begin{eqnarray}
\label{eomphi1ga}
&& - 18\, \phi_1 \varphi_0^{\prime}\, {}^{2} H^2 + 12\, \varphi_0^{\prime}\,  \varphi_1^{\prime}\,  H^2 + 18\, {\phi_1}^\prime\,  H\,  \varphi_0^{\prime}\, {}^{2}  \nonumber \\
&&+ 36\, \varphi_0^{\prime}\,  H\,  \varphi_0^{\prime \prime}\,  \phi_1  - 12\, \varphi_0^{\prime}\,  H\,  \varphi_1^{\prime \prime} - 12\, \varphi_1^{\prime}\,  H\,  \varphi_0^{\prime \prime} \nonumber \\
&& - 12\, \varphi_0^{\prime}\,  \varphi_1^{\prime}\,  a^{\prime \prime}\,  a{}^{-1} + 6\, {\delta}^{i j} H\,  {\partial}_{i j}{B_1}\,  \varphi_0^{\prime}\, {}^{2}  + 4\, {\delta}^{i j} \varphi_0^{\prime}\,  \varphi_0^{\prime \prime}\,  {\partial}_{i j}{B_1}\,  + \nonumber \\
&&4\, {\delta}^{i j} \varphi_0^{\prime}\,  H\,  {\partial}_{i j}{\varphi_1} + 4\, {\delta}^{i j} \varphi_0^{\prime \prime}\,  {\partial}_{i j}{\varphi_1}\,  + 2\, {\delta}^{i j} {\partial}_{i j}{B_1^\prime}\,  \varphi_0^{\prime}\, {}^{2} + V_{\varphi}\,  \phi_1 a{}^{4} + \nonumber \\
&&V_{\varphi \varphi}\,  a{}^{4} \varphi_1 + 2\, {\delta}^{i j} {\partial}_{i j}{\phi_1}\,  \varphi_0^{\prime}\, {}^{2}+ 18\, a^{\prime \prime}\,  \phi_1 \varphi_0^{\prime}\, {}^{2} a{}^{-1} = 0.
\end{eqnarray}

\section{Interaction Hamiltonian for higher order correlations of Galilean scalar field}\label{intHamGal}
Third order Interaction Hamiltonian of Galilean scalar field model
(\ref{action}), which is needed to compute Bi-spectrum, can be
obtained by substituting (\ref{fog}) in the Hamiltonian (\ref{fullH})
and extract the third order perturbed part as

\begin{eqnarray}
&& \mathcal{H}_{3} = \frac{1}{2}\, N_1 V_{\varphi \varphi} \varphi_1{}^{2} a{}^{3} + {N_1}^{i} \pi_{\lambda 1} {\partial}_{i}{\lambda_1}\,  + {N_1}^{i} \pi_{\varphi1} {\partial}_{i}{\varphi_1}\,  - {}\pi_{\lambda0} \pi_{\varphi1} N_1{}^{2} - \pi_{\lambda 1} {}\pi_{\varphi0} N_1{}^{2}   \nonumber \\
&&- 2\, N_0 N_1 \pi_{\lambda 1} \pi_{\varphi1} + \pi_{\lambda 1} \lambda_1 {\partial}_{i}{{N_1}^{i}}\,  -3\, N_1 S_0 N_0{}^{2} \pi_{\lambda 1}{}^{2} a{}^{3} - 6\, N_0 {}\pi_{\lambda0} \pi_{\lambda 1} S_0 N_1{}^{2} a{}^{3}  \nonumber \\
&&- S_0 N_1{}^{3} {}\pi_{\lambda0}{}^{2} a{}^{3} - S_1 N_0{}^{3} \pi_{\lambda 1}{}^{2} a{}^{3} -6\, N_1 {}\pi_{\lambda0} \pi_{\lambda 1} S_1 N_0{}^{2} a{}^{3} -3\, N_0 S_1 N_1{}^{2} {}\pi_{\lambda0}{}^{2} a{}^{3} \nonumber \\
&& - \frac{3}{2}\, N_0 {}\pi_{\lambda0} \pi_{\lambda 1} \kappa \lambda_1{}^{2} a{}^{(-3)} - \frac{3}{2}\, N_0 \kappa \lambda_0 \lambda_1 \pi_{\lambda 1}{}^{2} a{}^{(-3)} -\frac{3}{4}\, N_1 \kappa {}\pi_{\lambda0}{}^{2} \lambda_1{}^{2} a{}^{(-3)}  
 - \nonumber \\
&&{N_1}^{i} \pi_{\lambda 1} \lambda_0 {\partial}_{i}{N_1}\,  N_0{}^{-1} - {N_1}^{i} {}\pi_{\lambda0} \lambda_1 {\partial}_{i}{N_1}\,  N_0{}^{-1} - N_1 {\delta}^{i j} \lambda_0 {\partial}_{i}{N_1}\,  {\partial}_{j}{\varphi_1}\,  N_0{}^{(-2)} a{}^{(-2)} + \nonumber \\
&&~~~~~{\delta}^{i j} \lambda_1 {\partial}_{i}{N_1}\,  {\partial}_{j}{\varphi_1}\,  N_0{}^{-1} a{}^{(-2)} + N_1 S_0 {\delta}^{i j} {\partial}_{i}{\varphi_1}\,  {\partial}_{j}{\varphi_1}\,  a + N_0 S_1 {\delta}^{i j} {\partial}_{i}{\varphi_1}\,  {\partial}_{j}{\varphi_1}\,  a - \nonumber \\
&&~~~~~N_1 \pi_{\lambda 1} {\delta}_{i j} \kappa \lambda_0 {\pi_1}^{i j} a{}^{-1} - N_0 \pi_{\lambda 1} {\delta}_{i j} \kappa \lambda_1 {\pi_1}^{i j} a{}^{-1} - N_1 {}\pi_{\lambda0} {\delta}_{i j} \kappa \lambda_1 {\pi_1}^{i j} a{}^{-1}  - \nonumber \\
&&~~~~~N_1 {\delta}_{i j} {\delta}_{k l} \kappa {\pi_1}^{i j} {\pi_1}^{k l} a + 2\, N_1 {\delta}_{i j} {\delta}_{k l} \kappa {\pi_1}^{i k} {\pi_1}^{j l} a- \frac{3}{4}\, N_1 \kappa \pi_{\lambda 1}{}^{2} \lambda_0{}^{2} a{}^{(-3)} + \nonumber \\
&&N_1 {N_1}^{i} {}\pi_{\lambda0} \lambda_0 {\partial}_{i}{N_1}\,  N_0{}^{(-2)}- N_1 \pi_{\lambda 1} {\delta}_{i j} \kappa \lambda_1 \pi_0{}^{i j} a{}^{-1}- 3\, N_1 {}\pi_{\lambda0} \pi_{\lambda 1} \kappa \lambda_0 \lambda_1 a{}^{(-3)}.
\end{eqnarray}

First order and zeroth order Hamiltonian relations along with
(\ref{ncomp}) can be used to express the third order Hamiltonian in
terms of a single field and its derivatives so that we can compute the
correlation function.

Similarly, fourth Order interaction Hamiltonian, which helps to
compute Tri-spectrum, is given by
\begin{eqnarray}
&&\mathcal{H}_4 = - \pi_{\lambda 1} \pi_{\varphi1} N_1{}^{2} - 3\, N_0 S_0 N_1{}^{2} \pi_{\lambda 1}{}^{2} a{}^{3} - 2\, {}\pi_{\lambda0} \pi_{\lambda 1} S_0 N_1{}^{3} a{}^{3}   - \nonumber \\
&&~~~~S_1 N_1{}^{3} {}\pi_{\lambda0}{}^{2} a{}^{3} - \frac{3}{4}\, N_0 \kappa \pi_{\lambda 1}{}^{2} \lambda_1{}^{2} a{}^{(-3)} - \frac{3}{2}\, N_1 {}\pi_{\lambda0} \pi_{\lambda 1} \kappa \lambda_1{}^{2} a{}^{(-3)}  -  \nonumber \\
&&~~~~{N_1}^{i} {}\pi_{\lambda0} \lambda_0 {\partial}_{i}{N_1}\,  N_0{}^{(-3)} N_1{}^{2} +N_1 {N_1}^{i} \pi_{\lambda 1} \lambda_0 {\partial}_{i}{N_1}\,  N_0{}^{(-2)}  - \nonumber \\
&&~~~~{N_1}^{i} \pi_{\lambda 1} \lambda_1 {\partial}_{i}{N_1}\,  N_0{}^{-1} + {\delta}^{i j} \lambda_0 {\partial}_{i}{N_1}\,  {\partial}_{j}{\varphi_1}\,  N_0{}^{(-3)} N_1{}^{2} a{}^{(-2)}  +  \nonumber \\
&&~~~~N_1 S_1 {\delta}^{i j} {\partial}_{i}{\varphi_1}\,  {\partial}_{j}{\varphi_1}\,  a - N_1 \pi_{\lambda 1} {\delta}_{i j} \kappa \lambda_1 {\pi_1}^{i j} a{}^{-1}- 6\, N_0 {}\pi_{\lambda0} \pi_{\lambda 1} S_1 N_1{}^{2} a{}^{3}\nonumber \\
&&~~~~ -3\, N_1 S_1 N_0{}^{2} \pi_{\lambda 1}{}^{2} a{}^{3}+ N_1 {N_1}^{i} {}\pi_{\lambda0} \lambda_1 {\partial}_{i}{N_1}\,  N_0{}^{(-2)}- \frac{3}{2}\, N_1 \kappa \lambda_0 \lambda_1 \pi_{\lambda 1}{}^{2} a{}^{(-3)}\nonumber \\
&&~~~~- N_1 {\delta}^{i j} \lambda_1 {\partial}_{i}{N_1}\,  {\partial}_{j}{\varphi_1}\,  N_0{}^{(-2)} a{}^{(-2)}.
\end{eqnarray}

\section{Galilean and Canonical scalar field}\label{galcan}
Now we proceed to the action where both canonical part is present in a
Galilean fields model. The action is given by

\begin{eqnarray}
&&\mathcal{S} = \int d^4 x \sqrt{-g} \left[\frac{1}{2 \kappa} R - \frac{1}{2} \beta\, \partial_{\mu}{\varphi} \partial_{\nu}{\varphi} - \alpha\, g^{\mu \nu} \partial_{\mu}{\varphi} \partial_{\nu}{\varphi} \Box \varphi  - V(\varphi)\right] \nonumber \\
&&~~= \int d^4 x \sqrt{-g} \left[\frac{1}{2 \kappa} R - \frac{1}{2} \beta\, \partial_{\mu}{\varphi} \partial_{\nu}{\varphi} - \alpha\, g^{\mu \nu} \partial_{\mu}{\varphi} \partial_{\nu}{\varphi} S  - V(\varphi)\right] + \int d^4x \lambda \left( S - \Box \varphi \right).~~~~~~~~ \\
&&~~= \int \mathcal{L} \,d^4x 
\end{eqnarray}

Expanding the above action using ADM decomposition, we get {\small
\begin{eqnarray}
&& \mathcal{L} = S \lambda - N V \gamma{}^{\frac{1}{2}} + {\gamma}^{i j} {\partial}_{i}{\lambda}\,  {\partial}_{j}{\varphi}\,  + \lambda {\partial}_{i}{{\gamma}^{i j}}\,  {\partial}_{j}{\varphi}\,  - {\lambda^\prime}\,  {\varphi^\prime}\,  N{}^{(-2)} + \frac{1}{2}\, N ^{(3)}R \gamma{}^{\frac{1}{2}} \kappa{}^{-1} + {N}^{i} {\lambda^\prime}\,  {\partial}_{i}{\varphi}\,  N{}^{(-2)} + \nonumber \\
&&~~~~{N}^{i} {\varphi^\prime}\,  {\partial}_{i}{\lambda}\,  N{}^{(-2)} + \frac{1}{2}\, \beta N{}^{-1} \gamma{}^{\frac{1}{2}} {\varphi^\prime}\, {}^{2} + \lambda {N^\prime}\,  {\varphi^\prime}\,  N{}^{(-3)} - {N}^{i} {N}^{j} {\partial}_{i}{\lambda}\,  {\partial}_{j}{\varphi}\,  N{}^{(-2)} - {N}^{i} \lambda {N^\prime}\,  {\partial}_{i}{\varphi}\,  N{}^{(-3)} - \nonumber \\
&&~~~~{N}^{i} \lambda {\varphi^\prime}\,  {\partial}_{i}{N}\,  N{}^{(-3)} + S \alpha N{}^{-1} \gamma{}^{\frac{1}{2}} {\varphi^\prime}\, {}^{2} + {\gamma}^{i j} {\gamma}^{k l} \lambda {\partial}_{i}{{\gamma}_{j k}}\,  {\partial}_{l}{\varphi}\,  - \frac{1}{2}\, {\gamma}^{i j} {\gamma}^{k l} \lambda {\partial}_{i}{{\gamma}_{k l}}\,  {\partial}_{j}{\varphi}\,  + \nonumber \\
&&~~~~\frac{1}{2}\, {\gamma}^{i j} \lambda {{\gamma}_{i j}}{}^\prime\,  {\varphi^\prime}\,  N{}^{(-2)} - {\gamma}^{i j} \lambda {\partial}_{i}{N}\,  {\partial}_{j}{\varphi}\,  N{}^{-1} - \frac{1}{2}\, N \beta {\gamma}^{i j} {\partial}_{i}{\varphi}\,  {\partial}_{j}{\varphi}\,  \gamma{}^{\frac{1}{2}}%
 + {N}^{i} {N}^{j} \lambda {\partial}_{i}{N}\,  {\partial}_{j}{\varphi}\,  N{}^{(-3)} - \nonumber \\
&&~~~~{N}^{i} \beta {\varphi^\prime}\,  {\partial}_{i}{\varphi}\,  N{}^{-1} \gamma{}^{\frac{1}{2}} - \frac{1}{2}\, {N}^{i} {\gamma}^{j k} \lambda {{\gamma}_{j k}}{}^\prime\,  {\partial}_{i}{\varphi}\,  N{}^{(-2)} - \frac{1}{2}\, {N}^{i} {\gamma}^{j k} \lambda {\varphi^\prime}\,  {\partial}_{i}{{\gamma}_{j k}}\,  N{}^{(-2)} - \nonumber \\
&&~~~~\frac{1}{2}\, {K}^{i j} {K}^{k l} N {\gamma}_{i j} {\gamma}_{k l} \gamma{}^{\frac{1}{2}} \kappa{}^{-1} + \frac{1}{2}\, {K}^{i j} {K}^{k l} N {\gamma}_{i k} {\gamma}_{j l} \gamma{}^{\frac{1}{2}} \kappa{}^{-1} - N S \alpha {\gamma}^{i j} {\partial}_{i}{\varphi}\,  {\partial}_{j}{\varphi}\,  \gamma{}^{\frac{1}{2}} + \nonumber \\
&&~~~~\frac{1}{2}\, {N}^{i} {N}^{j} \beta {\partial}_{i}{\varphi}\,  {\partial}_{j}{\varphi}\,  N{}^{-1} \gamma{}^{\frac{1}{2}} + \frac{1}{2}\, {N}^{i} {N}^{j} {\gamma}^{k l} \lambda {\partial}_{i}{{\gamma}_{k l}}\,  {\partial}_{j}{\varphi}\,  N{}^{(-2)} - 2\, {N}^{i} S \alpha {\partial}_{0}{\varphi}\,  {\partial}_{i}{\varphi}\,  N{}^{-1} \gamma{}^{\frac{1}{2}} + \nonumber \\
&&~~~~{N}^{i} {N}^{j} S \alpha {\partial}_{i}{\varphi}\,  {\partial}_{j}{\varphi}\,  N{}^{-1} \gamma{}^{\frac{1}{2}}
\end{eqnarray}}

Definition of all momenta are same as we derived in the Galilean case except $\pi_\varphi$ and it is given by
\begin{eqnarray}
&& \pi_\varphi = - {\lambda^\prime}\,  N{}^{(-2)} + {N}^{i} {\partial}_{i}{\lambda}\,  N{}^{(-2)} + \beta N{}^{-1} \gamma{}^{\frac{1}{2}} {\varphi^\prime}\,  + \lambda {N^\prime}\,  N{}^{(-3)} - {N}^{i} \lambda {\partial}_{i}{N}\,  N{}^{(-3)} + \nonumber \\
&&~~~~ 2\, S \alpha N{}^{-1} \gamma{}^{\frac{1}{2}} {\varphi^\prime}\,  + \frac{1}{2}\, {\gamma}^{i j} \lambda {{\gamma}_{i j}}{}^\prime\,  N{}^{(-2)} - {N}^{i} \beta {\partial}_{i}{\varphi}\,  N{}^{-1} \gamma{}^{\frac{1}{2}} - \frac{1}{2}\, {N}^{i} {\gamma}^{j k} \lambda {\partial}_{i}{{\gamma}_{j k}}\,  N{}^{(-2)} - \nonumber \\
&&~~~~2\, {N}^{i} S \alpha {\partial}_{i}{\varphi}\,  N{}^{-1} \gamma{}^{\frac{1}{2}}.
\end{eqnarray}

Then the Dirac-Hamiltonian becomes,
\begin{eqnarray}
&&  \mathcal{H}_D = - S \lambda + N V \gamma{}^{\frac{1}{2}} + {N}^{i} \pi_\lambda {\partial}_{i}{\lambda}\,  + {N}^{i} \pi_\varphi {\partial}_{i}{\varphi}\,  - \pi_\lambda \pi_\varphi N{}^{2} + \pi_\lambda \lambda {\partial}_{i}{{N}^{i}}\,  +  2 {\gamma}_{i j} {\partial}_{k}{{N}^{i}}\,  {\pi}^{j k} -  \nonumber \\
&&~~~~~~ {\gamma}^{i j} {\partial}_{i}{\lambda}\,  {\partial}_{j}{\varphi}\,  - \lambda {\partial}_{i}{{\gamma}^{i j}}\,  {\partial}_{j}{\varphi}\,  - \frac{1}{2}\, N ^{(3)}R \gamma{}^{\frac{1}{2}} \kappa{}^{-1} - \frac{1}{2}\, \beta N{}^{3} \pi_\lambda{}^{2} \gamma{}^{\frac{1}{2}} - \frac{3}{4}\, N \kappa \pi_\lambda{}^{2} \gamma{}^{-\frac{1}{2}} \lambda{}^{2} -\nonumber \\
&&~~~~~~ {N}^{i} \pi_\lambda \lambda {\partial}_{i}{N}\,  N{}^{-1} +  {N}^{i}  {\partial}_{i}{{\gamma}_{l m}}\,  {\pi}^{l m} -  {N}^{i}  {\partial}_{l}{{\gamma}_{i m}}\,  {\pi}^{l m} + {N}^{i}  {\partial}_{m}{{\gamma}_{i l}}\,  {\pi}^{l m}  - S \alpha N{}^{3} \pi_\lambda{}^{2} \gamma{}^{\frac{1}{2}} -  \nonumber \\
&&~~~~~~{\gamma}^{i j} {\gamma}^{k l} \lambda {\partial}_{i}{{\gamma}_{j k}}\,  {\partial}_{l}{\varphi}\,  +\frac{1}{2}\, {\gamma}^{i j} {\gamma}^{k l} \lambda {\partial}_{i}{{\gamma}_{k l}}\,  {\partial}_{j}{\varphi}\,  + {\gamma}^{i j} \lambda {\partial}_{i}{N}\,  {\partial}_{j}{\varphi}\,  N{}^{-1} + \frac{1}{2}\, N \beta {\gamma}^{i j} {\partial}_{i}{\varphi}\,  {\partial}_{j}{\varphi}\,  \gamma{}^{\frac{1}{2}} -  \nonumber \\
&&~~~~~~ N \pi_\lambda {\gamma}_{i j} \kappa \lambda {\pi}^{i j} \gamma{}^{-\frac{1}{2}} + N S \alpha {\gamma}^{i j} {\partial}_{i}{\varphi}\,  {\partial}_{j}{\varphi}\,  \gamma{}^{\frac{1}{2}} -N {\gamma}_{i j} {\gamma}_{k l} \kappa {\pi}^{i j} {\pi}^{k l} \gamma{}^{-\frac{1}{2})} + 2\, N {\gamma}_{i j} {\gamma}_{k l} \kappa {\pi}^{i k} {\pi}^{j l} \gamma{}^{-\frac{1}{2}}  \nonumber \\
&&~~~~~~ + \xi (\pi_N + \lambda N{}^{-1} \pi_\lambda)
\end{eqnarray}

\subsection{Zeroth order}
Zeroth order Hamiltonian, in terms of $\pi_a$, takes the form
\begin{eqnarray}
&& \mathcal{H}_{D0} = - S_0 \lambda_0 + N_0 V a{}^{3} - {}\pi_{\lambda0} {}\pi_{\varphi0} N_0{}^{2} - \frac{1}{2}\, \beta N_0{}^{3} {}\pi_{\lambda0}{}^{2} a{}^{3} - \frac{3}{4}\, N_0 \kappa {}\pi_{\lambda0}{}^{2} \lambda_0{}^{2} a{}^{(-3)} \nonumber \\
&& - S_0 \alpha N_0{}^{3} {}\pi_{\lambda0}{}^{2} a{}^{3}- \frac{1}{2}\, N_0 \pi_a {}\pi_{\lambda0} \kappa \lambda_0 a{}^{(-2)} - \frac{1}{12}\, N_0 \kappa \pi_a{}^{2} a{}^{-1} + \xi_0 \left( \pi_{N0} + {}\pi_{\lambda0} \lambda_0 N_0{}^{-1} \right).~~~~~~
\end{eqnarray}

Zeroth order Hamiltonian constraint or the equation of motion of $N_0$
is given by
\begin{equation}
V a{}^{2} + \frac{1}{2}\, \beta  \varphi_0^{\prime}\, {}^{2} - 3\,  \kappa{}^{-1} H^2 - 6\, \alpha H\,  a{}^{(-2)} \varphi_0^{\prime}\, {}^{3} = 0.
\end{equation}

Similarly, equations of motion of $a$ and $\varphi_0$ are given by
 \begin{eqnarray}
&& 6\, \alpha H\,  \varphi_0^{\prime}\, {}^{3} a{}^{(-2)} - 6\, \alpha \varphi_0^{\prime \prime}\,  \varphi_0^{\prime}\, {}^{2} a{}^{-2} + 3\, \kappa{}^{-1} H^2 - \nonumber \\
&&~~~~~~~~6\, a^{\prime \prime}\,a^{-1}  \kappa{}^{-1} + 3\, V a{}^{2} - \frac{3}{2}\, \beta \varphi_0^{\prime}\, {}^{2}  = 0, \\ 
&& V_\varphi\,  a{}^{2}- 6\, \alpha a^{\prime \prime}\,  \varphi_0^{\prime}\, {}^{2} a{}^{(-3)} + 6\, \alpha \varphi_0^{\prime}\, {}^{2} H^2 a{}^{(-2)} - 12\, \alpha \varphi_0^{\prime}\,  H\,  \varphi_0^{\prime \prime}\,  a{}^{(-2)} \nonumber \\
&&~~~~+ 2\, \beta \varphi_0^{\prime}\,  H + \beta \varphi_0^{\prime \prime}\, = 0.
 \end{eqnarray}

\subsection{First order}
first order Hamilton's equations are obtained from second order
perturbed Hamiltonian and it is given by
\begin{eqnarray}
&& \mathcal{H}_{D2} = - S_1 \lambda_1 + N_1 V_\varphi\,  a{}^{3} \varphi_1 + \frac{1}{2}\, N_0 V_{\varphi \varphi}\,  \varphi_1{}^{2} a{}^{3} + {N_1}^{i} {}\pi_{\lambda0} {\partial}_{i}{\lambda_1}\,  + {N_1}^{i} {}\pi_{\varphi0} {\partial}_{i}{\varphi_1}\,   - \nonumber \\
&&~~~~2\, N_0 N_1 {}\pi_{\lambda0} \pi_{\varphi1} - 2\, N_0 N_1 \pi_{\lambda 1} {}\pi_{\varphi0} - \pi_{\lambda 1} \pi_{\varphi1} N_0{}^{2} + \pi_{\lambda 1} \lambda_0 {\partial}_{i}{{N_1}^{i}}\,   -\nonumber \\
&&~~~~  {\delta}^{i j} {\partial}_{i}{\lambda_1}\,  {\partial}_{j}{\varphi_1}\,  a{}^{(-2)} - \frac{1}{2}\, \beta N_0{}^{3} \pi_{\lambda 1}{}^{2} a{}^{3} - 3\, N_1 {}\pi_{\lambda0} \pi_{\lambda 1} \beta N_0{}^{2} a{}^{3} - \frac{3}{2}\, N_0 \beta N_1{}^{2} {}\pi_{\lambda0}{}^{2} a{}^{3} -\nonumber \\
&&~~~~ \frac{3}{4}\, N_0 \kappa {}\pi_{\lambda0}{}^{2} \lambda_1{}^{2} a{}^{(-3)} - 3\, N_0 {}\pi_{\lambda0} \pi_{\lambda 1} \kappa \lambda_0 \lambda_1 a{}^{(-3)}%
 - \frac{3}{4}\, N_0 \kappa \pi_{\lambda 1}{}^{2} \lambda_0{}^{2} a{}^{(-3)}  -\nonumber \\
&&~~~~ \frac{3}{2}\, N_1 {}\pi_{\lambda0} \pi_{\lambda 1} \kappa \lambda_0{}^{2} a{}^{(-3)} - {N_1}^{i} {}\pi_{\lambda0} \lambda_0 {\partial}_{i}{N_1}\,  N_0{}^{-1} - S_0 \alpha N_0{}^{3} \pi_{\lambda 1}{}^{2} a{}^{3}  - \nonumber \\
&&~~~~ 3\, N_0 S_0 \alpha N_1{}^{2} {}\pi_{\lambda0}{}^{2} a{}^{3} - 2\, {}\pi_{\lambda0} \pi_{\lambda 1} S_1 \alpha N_0{}^{3} a{}^{3} - 3\, N_1 S_1 \alpha N_0{}^{2} {}\pi_{\lambda0}{}^{2} a{}^{3}  + \nonumber \\
&&~~~~\frac{1}{2}\, N_0 \beta {\delta}^{i j} {\partial}_{i}{\varphi_1}\,  {\partial}_{j}{\varphi_1}\,  a - N_0 \pi_{\lambda 1} {\delta}_{i j} \kappa \lambda_0 {\pi_1}^{i j} a{}^{-1} - N_1 {}\pi_{\lambda0} {\delta}_{i j} \kappa \lambda_0 {\pi_1}^{i j} a{}^{-1}  -\nonumber \\
&&~~~~ N_0 {}\pi_{\lambda0} {\delta}_{i j} \kappa \lambda_1 {\pi_1}^{i j} a{}^{-1} - N_0 \pi_{\lambda 1} {\delta}_{i j} \kappa \lambda_1 \pi_0{}^{i j} a{}^{-1} - N_1 {}\pi_{\lambda0} {\delta}_{i j} \kappa \lambda_1 \pi_0{}^{i j} a{}^{-1}  -\nonumber \\
&&~~~~ N_0 {\delta}_{i j} {\delta}_{k l} \kappa {\pi_1}^{i j} {\pi_1}^{k l} a - 2\, N_1 {\delta}_{i j} {\delta}_{k l} \kappa \pi_0{}^{i j} {\pi_1}^{k l} a%
 + 2\, N_0 {\delta}_{i j} {\delta}_{k l} \kappa {\pi_1}^{i k} {\pi_1}^{j l} a   +\nonumber \\
&&~~~~  \pi_{N1} \xi_1 - N_1 {}\pi_{\lambda0} \lambda_0 N_0{}^{(-2)} \xi_1 + {}\pi_{\lambda0} \lambda_0 N_0{}^{(-3)} N_1{}^{2} \xi_0 + \pi_{\lambda 1} \lambda_0 N_0{}^{-1} \xi_1  + \nonumber \\
&&~~~~ {}\pi_{\lambda0} \lambda_1 N_0{}^{-1} \xi_1 - N_1 {}\pi_{\lambda0} \lambda_1 N_0{}^{(-2)} \xi_0 + \pi_{\lambda 1} \lambda_1 N_0{}^{-1} \xi_0- {}\pi_{\lambda0} {}\pi_{\varphi0} N_1{}^{2} +\nonumber \\
&&~~~~ {}\pi_{\lambda0} \lambda_1 {\partial}_{i}{{N_1}^{i}}\,  +  2  {\delta}_{i j} {\partial}_{k}{{N_1}^{i}}\,  {\pi_1}^{j k} a{}^{2}- \frac{3}{2}\, N_1 \kappa \lambda_0 \lambda_1 {}\pi_{\lambda0}{}^{2} a{}^{(-3)}-\nonumber \\
&&~~~~6\, N_1 {}\pi_{\lambda0} \pi_{\lambda 1} S_0 \alpha N_0{}^{2} a{}^{3} + {\delta}^{i j} \lambda_0 {\partial}_{i}{N_1}\,  {\partial}_{j}{\varphi_1}\,  N_0{}^{-1} a{}^{(-2)}- N_1 \pi_{\lambda 1} {\delta}_{i j} \kappa \lambda_0 \pi_0{}^{i j} a{}^{-1}+\nonumber \\
&&~~~~ N_0 S_0 \alpha {\delta}^{i j} {\partial}_{i}{\varphi_1}\,  {\partial}_{j}{\varphi_1}\,  a- N_1 \pi_{\lambda 1} \lambda_0 N_0{}^{(-2)} \xi_0+ 4\, N_1 {\delta}_{i j} {\delta}_{k l} \kappa {{}\pi_0{}}^{i k} {\pi_1}^{j l} a
\end{eqnarray}
Since we have got the perturbed second order Hamiltonian, we can
obtained the field equations using Hamilton's equations. First order
Momentum constraint or the equation of motion of $N_1^{i}$ is given by
\begin{eqnarray}
 &&- 6\, \alpha H\,  {\partial}_{i}{\varphi_1}\,  \varphi_0^{\prime}\, {}^{2}  + \beta \varphi_0^{\prime}\,  {\partial}_{i}{\varphi_1}\,  a{}^{2} - 2\, \alpha {\partial}_{i}{\phi_1}\,  \varphi_0^{\prime}\, {}^{3} +  2\, \alpha {\partial}_{i}{\varphi_1^\prime}\,  \varphi_0^{\prime}\, {}^{2} - \nonumber \\
 &&~~~~~~~~2\, H\,  {\partial}_{i}{\phi_1}\,  \kappa{}^{-1} a^2 = 0.
\end{eqnarray}

Similarly, First order Hamiltonian constraint or the equation of
motion of $N_1$ is given by
\begin{eqnarray}
&&24\, \alpha H\,  \phi_1 \varphi_0^{\prime}\, {}^{3} a{}^{(-2)} - 18\, \alpha \varphi_1^{\prime}\,  H\,  \varphi_0^{\prime}\, {}^{2} a{}^{(-2)} + V_\varphi \,  a{}^{2} \varphi_1 - \beta \phi_1 \varphi_0^{\prime}\, {}^{2}  +\nonumber \\
 && 2\, \alpha {\delta}^{i j} {\partial}_{i j}{B_1}\,  \varphi_0^{\prime}\, {}^{3} a{}^{-2} + \beta \varphi_0^{\prime}\,  \varphi_1^{\prime} + 2\, \alpha {\delta}^{i j} {\partial}_{i j}{\varphi_1}\,  \varphi_0^{\prime}\, {}^{2} a{}^{-2} \nonumber \\
 &&+ 2\, {\delta}^{i j} H\,  {\partial}_{i j}{B_1}\,  \kappa{}^{-1} + 6\, \phi_1 \kappa{}^{-1} H = 0
\end{eqnarray}
and the equation of motion of $\varphi_1$ is given by

\begin{eqnarray}
&&- 18\, \alpha \phi_1 \varphi_0^{\prime}\, {}^{2} H^2 + 12\, \alpha \varphi_0^{\prime}\,  \varphi_1^{\prime}\,  H^2 + 18\, \alpha {\phi_1^\prime}\,  H\,  \varphi_0^{\prime}\,{}^{2} \nonumber \\
 && + 36\, \alpha \varphi_0^{\prime}\,  H\,  \varphi_0^{\prime \prime}\,  \phi_1  - 12\, \alpha \varphi_0^{\prime}\,  H\,  \varphi_1^{\prime \prime} - 12\, \alpha \varphi_1^{\prime}\,H\,  \varphi_0^{\prime \prime}\nonumber \\
 && + 18\, \alpha a^{\prime \prime}\,  \phi_1 \varphi_0^{\prime}\, {}^{2} a{}^{-1} - 12\, \alpha \varphi_0^{\prime}\,  \varphi_1^{\prime}\,  a^{\prime \prime}\,  a{}^{-1} - 2\, \beta \varphi_0^{\prime}\,  H\,  \phi_1 a^2 - \beta {\phi_1^\prime}\,  \varphi_0^{\prime}\,  a{}^{2}\nonumber \\
 && - \beta \varphi_0^{\prime \prime}\,  \phi_1 a{}^{2} + 6\, \alpha {\delta}^{i j} H\,  {\partial}_{i j}{B_1}\,  \varphi_0^{\prime}\, {}^{2}  + 4\, \alpha {\delta}^{i j} \varphi_0^{\prime}\,  \varphi_0^{\prime \prime}\,  {\partial}_{i j}{B_1}\,  + 2\, \beta \varphi_1^{\prime}\,  H\,  a^2\nonumber \\
 && + \beta \varphi_1^{\prime \prime}\,  a{}^{2} + 4\, \alpha {\delta}^{i j} \varphi_0^{\prime}\,  H\,  {\partial}_{i j}{\varphi_1} + 4\, \alpha {\delta}^{i j} \varphi_0^{\prime \prime}\,  {\partial}_{i j}{\varphi_1}\,  + 2\, \alpha {\delta}^{i j} {\partial}_{i j}{B_1^\prime}\,  \varphi_0^{\prime}\, {}^{2}\nonumber \\
 && + V_\varphi\,  \phi_1 a{}^{4}%
 + V_{\varphi \varphi}\,  a{}^{4} \varphi_1 - \beta {\delta}^{i j} \varphi_0^{\prime}\,  {\partial}_{i j}{B_1}\,  a{}^{2} + 2\, \alpha {\delta}^{i j} {\partial}_{i j}{\phi_1}\,  \varphi_0^{\prime}\, {}^{2} - \beta {\delta}^{i j} {\partial}_{i j}{\varphi_1}\,  a{}^{2} = 0.~~~~~~~~~
\end{eqnarray}


\begin{thebibliography}{10}

\bibitem{Bardeen:1980kt}
J.~M. Bardeen, {\it {Gauge Invariant Cosmological Perturbations}},  {\em
  Phys.Rev.} {\bf D22} (1980) 1882--1905.

\bibitem{Langlois1994}
D.~{Langlois}, {\it {Hamiltonian formalism and gauge invariance for linear
  perturbations in inflation}},  {\em Classical and Quantum Gravity} {\bf 11}
  (1994) 389--407.

\bibitem{Mukhanov:1990me}
V.~F. Mukhanov, H.~Feldman, and R.~H. Brandenberger, {\it {Theory of
  cosmological perturbations. Part 1. Classical perturbations. Part 2. Quantum
  theory of perturbations. Part 3. Extensions}},  {\em Phys.Rept.} {\bf 215}
  (1992) 203--333.

\bibitem{Sasaki1986}
M.~Sasaki, {\it {Large Scale Quantum Fluctuations in the Inflationary
  Universe}},  {\em Progress of Theoretical Physics} {\bf 76} (1986)
  1036--1046.

\bibitem{Lidsey:1995np}
J.~E. Lidsey, A.~R. Liddle, E.~W. Kolb, E.~J. Copeland, T.~Barreiro, {\em
  et.~al.}, {\it {Reconstructing the inflation potential : An overview}},  {\em
  Rev.Mod.Phys.} {\bf 69} (1997) 373--410,
  [\href{http://xxx.lanl.gov/abs/astro-ph/9508078}{{\tt astro-ph/9508078}}].

\bibitem{Garriga1999}
J.~Garriga and V.~F. Mukhanov, {\it {Perturbations in k-inflation}},  {\em
  Phys.Lett.} {\bf B458} (1999) 219--225,
  [\href{http://xxx.lanl.gov/abs/hep-th/9904176}{{\tt hep-th/9904176}}].

\bibitem{Kodama01011984}
H.~Kodama and M.~Sasaki, {\it Cosmological perturbation theory},  {\em Progress
  of Theoretical Physics Supplement} {\bf 78} (1984) 1--166,
  [\href{http://xxx.lanl.gov/abs/http://ptps.oxfordjournals.org/content/78/1.full.pdf+html}{{\tt
  http://ptps.oxfordjournals.org/content/78/1.full.pdf+html}}].

\bibitem{Komatsu:2008hk}
{\bf WMAP Collaboration} Collaboration, E.~Komatsu {\em et.~al.}, {\it
  {Five-Year Wilkinson Microwave Anisotropy Probe (WMAP) Observations:
  Cosmological Interpretation}},  {\em Astrophys.J.Suppl.} {\bf 180} (2009)
  330--376, [\href{http://xxx.lanl.gov/abs/0803.0547}{{\tt arXiv:0803.0547}}].

\bibitem{STAROBINSKY1982175}
A.~Starobinsky, {\it Dynamics of phase transition in the new inflationary
  universe scenario and generation of perturbations},  {\em Physics Letters B}
  {\bf 117} (1982), no.~3 175 -- 178.

\bibitem{BardeenSteinhardt1983}
J.~M. Bardeen, P.~J. Steinhardt, and M.~S. Turner, {\it Spontaneous creation of
  almost scale-free density perturbations in an inflationary universe},  {\em
  Phys. Rev. D} {\bf 28} (1983) 679--693.

\bibitem{Salopek1990}
D.~S. Salopek and J.~R. Bond, {\it {Nonlinear evolution of long-wavelength
  metric fluctuations in inflationary models}},  {\em Physical Review D} {\bf
  42} (1990) 3936--3962.

\bibitem{Sasaki:1995aw}
M.~Sasaki and E.~D. Stewart, {\it {A General analytic formula for the spectral
  index of the density perturbations produced during inflation}},  {\em Prog.
  Theor. Phys.} {\bf 95} (1996) 71--78,
  [\href{http://xxx.lanl.gov/abs/astro-ph/9507001}{{\tt astro-ph/9507001}}].

\bibitem{Lyth:1998xn}
D.~H. Lyth and A.~Riotto, {\it {Particle physics models of inflation and the
  cosmological density perturbation}},  {\em Phys.Rept.} {\bf 314} (1999)
  1--146, [\href{http://xxx.lanl.gov/abs/hep-ph/9807278}{{\tt
  hep-ph/9807278}}].

\bibitem{Armendariz-Picon1999}
C.~Armendariz-Picon, T.~Damour, and V.~F. Mukhanov, {\it {k - inflation}},
  {\em Phys.Lett.} {\bf B458} (1999) 209--218,
  [\href{http://xxx.lanl.gov/abs/hep-th/9904075}{{\tt hep-th/9904075}}].

\bibitem{Lyth:2007qh}
D.~H. Lyth, {\it {Particle physics models of inflation}},  {\em Lect. Notes
  Phys.} {\bf 738} (2008) 81--118,
  [\href{http://xxx.lanl.gov/abs/hep-th/0702128}{{\tt hep-th/0702128}}].

\bibitem{Mazumdar:2010sa}
A.~Mazumdar and J.~Rocher, {\it {Particle physics models of inflation and
  curvaton scenarios}},  {\em Phys. Rept.} {\bf 497} (2011) 85--215,
  [\href{http://xxx.lanl.gov/abs/1001.0993}{{\tt arXiv:1001.0993}}].

\bibitem{Bruni:1996im}
M.~Bruni, S.~Matarrese, S.~Mollerach, and S.~Sonego, {\it {Perturbations of
  space-time: Gauge transformations and gauge invariance at second order and
  beyond}},  {\em Class.Quant.Grav.} {\bf 14} (1997) 2585--2606,
  [\href{http://xxx.lanl.gov/abs/gr-qc/9609040}{{\tt gr-qc/9609040}}].

\bibitem{Acquaviva:2002ud}
V.~Acquaviva, N.~Bartolo, S.~Matarrese, and A.~Riotto, {\it {Second order
  cosmological perturbations from inflation}},  {\em Nucl. Phys.} {\bf B667}
  (2003) 119--148, [\href{http://xxx.lanl.gov/abs/astro-ph/0209156}{{\tt
  astro-ph/0209156}}].

\bibitem{Nakamura:2003wk}
K.~Nakamura, {\it {Gauge invariant variables in two parameter nonlinear
  perturbations}},  {\em Prog. Theor. Phys.} {\bf 110} (2003) 723--755,
  [\href{http://xxx.lanl.gov/abs/gr-qc/0303090}{{\tt gr-qc/0303090}}].

\bibitem{PhysRevD.69.104011}
H.~Noh and J.-c. Hwang, {\it Second-order perturbations of the friedmann world
  model},  {\em Phys. Rev. D} {\bf 69} (May, 2004) 104011.

\bibitem{Bartolo:2001cw}
N.~Bartolo, S.~Matarrese, and A.~Riotto, {\it {Nongaussianity from inflation}},
   {\em Phys. Rev.} {\bf D65} (2002) 103505,
  [\href{http://xxx.lanl.gov/abs/hep-ph/0112261}{{\tt hep-ph/0112261}}].

\bibitem{Rigopoulos:2002mc}
G.~Rigopoulos, {\it {On second order gauge invariant perturbations in
  multi-field inflationary models}},  {\em Class. Quant. Grav.} {\bf 21} (2004)
  1737--1754, [\href{http://xxx.lanl.gov/abs/astro-ph/0212141}{{\tt
  astro-ph/0212141}}].

\bibitem{Bernardeau:2002jy}
F.~Bernardeau and J.-P. Uzan, {\it {NonGaussianity in multifield inflation}},
  {\em Phys. Rev.} {\bf D66} (2002) 103506,
  [\href{http://xxx.lanl.gov/abs/hep-ph/0207295}{{\tt hep-ph/0207295}}].

\bibitem{Bernardeau:2002jf}
F.~Bernardeau and J.-P. Uzan, {\it {Inflationary models inducing non-Gaussian
  metric fluctuations}},  {\em Phys. Rev.} {\bf D67} (2003) 121301,
  [\href{http://xxx.lanl.gov/abs/astro-ph/0209330}{{\tt astro-ph/0209330}}].

\bibitem{Malik:2003mv}
K.~A. Malik and D.~Wands, {\it {Evolution of second-order cosmological
  perturbations}},  {\em Class. Quant. Grav.} {\bf 21} (2004) L65--L72,
  [\href{http://xxx.lanl.gov/abs/astro-ph/0307055}{{\tt astro-ph/0307055}}].

\bibitem{Bartolo:2003bz}
N.~Bartolo, S.~Matarrese, and A.~Riotto, {\it {Evolution of second - order
  cosmological perturbations and non-Gaussianity}},  {\em JCAP} {\bf 0401}
  (2004) 003, [\href{http://xxx.lanl.gov/abs/astro-ph/0309692}{{\tt
  astro-ph/0309692}}].

\bibitem{Finelli:2003bp}
F.~Finelli, G.~Marozzi, G.~P. Vacca, and G.~Venturi, {\it {Energy momentum
  tensor of cosmological fluctuations during inflation}},  {\em Phys. Rev.}
  {\bf D69} (2004) 123508, [\href{http://xxx.lanl.gov/abs/gr-qc/0310086}{{\tt
  gr-qc/0310086}}].

\bibitem{Bartolo:2004if}
N.~Bartolo, E.~Komatsu, S.~Matarrese, and A.~Riotto, {\it {Non-Gaussianity from
  inflation: Theory and observations}},  {\em Phys. Rept.} {\bf 402} (2004)
  103--266, [\href{http://xxx.lanl.gov/abs/astro-ph/0406398}{{\tt
  astro-ph/0406398}}].

\bibitem{Enqvist:2004bk}
K.~Enqvist and A.~Vaihkonen, {\it {Non-Gaussian perturbations in hybrid
  inflation}},  {\em JCAP} {\bf 0409} (2004) 006,
  [\href{http://xxx.lanl.gov/abs/hep-ph/0405103}{{\tt hep-ph/0405103}}].

\bibitem{Vernizzi:2004nc}
F.~Vernizzi, {\it {On the conservation of second-order cosmological
  perturbations in a scalar field dominated Universe}},  {\em Phys. Rev.} {\bf
  D71} (2005) 061301, [\href{http://xxx.lanl.gov/abs/astro-ph/0411463}{{\tt
  astro-ph/0411463}}].

\bibitem{Tomita:2005et}
K.~Tomita, {\it {Relativistic second-order perturbations of nonzero-lambda flat
  cosmological models and CMB anisotropies}},  {\em Phys. Rev.} {\bf D71}
  (2005) 083504, [\href{http://xxx.lanl.gov/abs/astro-ph/0501663}{{\tt
  astro-ph/0501663}}].

\bibitem{Lyth:2005du}
D.~H. Lyth and Y.~Rodriguez, {\it {Non-Gaussianity from the second-order
  cosmological perturbation}},  {\em Phys. Rev.} {\bf D71} (2005) 123508,
  [\href{http://xxx.lanl.gov/abs/astro-ph/0502578}{{\tt astro-ph/0502578}}].

\bibitem{Seery:2005gb}
D.~Seery and J.~E. Lidsey, {\it {Primordial non-Gaussianities from
  multiple-field inflation}},  {\em JCAP} {\bf 0509} (2005) 011,
  [\href{http://xxx.lanl.gov/abs/astro-ph/0506056}{{\tt astro-ph/0506056}}].

\bibitem{Malik:2005cy}
K.~A. Malik, {\it {Gauge-invariant perturbations at second order: Multiple
  scalar fields on large scales}},  {\em JCAP} {\bf 0511} (2005) 005,
  [\href{http://xxx.lanl.gov/abs/astro-ph/0506532}{{\tt astro-ph/0506532}}].

\bibitem{Seery:2006vu}
D.~Seery, J.~E. Lidsey, and M.~S. Sloth, {\it {The inflationary trispectrum}},
  {\em JCAP} {\bf 0701} (2007) 027,
  [\href{http://xxx.lanl.gov/abs/astro-ph/0610210}{{\tt astro-ph/0610210}}].

\bibitem{Lifshitz:1963ps}
E.~Lifshitz and I.~Khalatnikov, {\it {Investigations in relativistic
  cosmology}},  {\em Adv.Phys.} {\bf 12} (1963) 185--249.

\bibitem{Lukash:1980iv}
V.~Lukash, {\it {Production of phonons in an isotropic universe}},  {\em
  Sov.Phys.JETP} {\bf 52} (1980) 807--814.

\bibitem{Ostro}
M.~Ostrogradsky, {\it {Memoires sur les equations differentielles relatives au
  probleme des isoperimetres}},  {\em Mem. Ac. St. Petersbourg} {\bf VI} (1850)
  385.

\bibitem{Woodard:2015zca}
R.~P. Woodard, {\it {The Theorem of Ostrogradsky}},
  \href{http://xxx.lanl.gov/abs/1506.0221}{{\tt arXiv:1506.0221}}.

\bibitem{Kobayashi2010}
T.~Kobayashi, M.~Yamaguchi, and J.~Yokoyama, {\it {G-inflation: Inflation
  driven by the Galileon field}},  {\em Phys.Rev.Lett.} {\bf 105} (2010)
  231302, [\href{http://xxx.lanl.gov/abs/1008.0603}{{\tt arXiv:1008.0603}}].

\bibitem{Kobayashi2011}
T.~Kobayashi, M.~Yamaguchi, and J.~Yokoyama, {\it {Generalized G-inflation:
  Inflation with the most general second-order field equations}},  {\em
  Prog.Theor.Phys.} {\bf 126} (2011) 511--529,
  [\href{http://xxx.lanl.gov/abs/1105.5723}{{\tt arXiv:1105.5723}}].

\bibitem{Deffayet2009}
C.~Deffayet, G.~Esposito-Farese, and A.~Vikman, {\it {Covariant Galileon}},
  {\em Phys.Rev.} {\bf D79} (2009) 084003,
  [\href{http://xxx.lanl.gov/abs/0901.1314}{{\tt arXiv:0901.1314}}].

\bibitem{Huang:2006eha}
X.~Chen, M.-x. Huang, and G.~Shiu, {\it {The Inflationary Trispectrum for
  Models with Large Non-Gaussianities}},  {\em Phys. Rev.} {\bf D74} (2006)
  121301, [\href{http://xxx.lanl.gov/abs/hep-th/0610235}{{\tt
  hep-th/0610235}}].

\bibitem{Deffayet:2015qwa}
C.~Deffayet, G.~Esposito-Farese, and D.~A. Steer, {\it {Counting the degrees of
  freedom of generalized Galileons}},  {\em Phys. Rev.} {\bf D92} (2015)
  084013, [\href{http://xxx.lanl.gov/abs/1506.0197}{{\tt arXiv:1506.0197}}].

\bibitem{Barrau:2013ula}
A.~Barrau, T.~Cailleteau, J.~Grain, and J.~Mielczarek, {\it {Observational
  issues in loop quantum cosmology}},  {\em Class. Quant. Grav.} {\bf 31}
  (2014) 053001, [\href{http://xxx.lanl.gov/abs/1309.6896}{{\tt
  arXiv:1309.6896}}].

\bibitem{Ashtekar:2013xka}
A.~Ashtekar, {\it {Loop Quantum Gravity and the The Planck Regime of
  Cosmology}},  in {\em {Proceedings, Relativity and Gravitation : 100 Years
  after Einstein in Prague}}, 2013.
\newblock \href{http://xxx.lanl.gov/abs/1303.4989}{{\tt arXiv:1303.4989}}.

\bibitem{Bojowald:2012xy}
M.~Bojowald, {\it {Quantum Cosmology: Effective Theory}},  {\em Class. Quant.
  Grav.} {\bf 29} (2012) 213001, [\href{http://xxx.lanl.gov/abs/1209.3403}{{\tt
  arXiv:1209.3403}}].

\bibitem{Nandi:2015tha}
D.~Nandi and S.~Shankaranarayanan, {\it {`Constraint consistency' at all orders
  in cosmological perturbation theory}},  {\em JCAP} {\bf 1508} (2015), no.~08
  050, [\href{http://xxx.lanl.gov/abs/1502.04036}{{\tt arXiv:1502.04036}}].

\bibitem{Andrzejewski:2010kz}
K.~Andrzejewski, J.~Gonera, P.~Machalski, and P.~Maslanka, {\it {Modified
  Hamiltonian formalism for higher-derivative theories}},  {\em Phys. Rev.}
  {\bf D82} (2010) 045008, [\href{http://xxx.lanl.gov/abs/1005.3941}{{\tt
  arXiv:1005.3941}}].

\bibitem{Nesterenko}
V.~V. Nesterenko, {\it Singular lagrangians with higher derivatives},  {\em
  Journal of Physics A: Mathematical and General} {\bf 22} (1989), no.~10 1673.

\bibitem{Chen:2012au}
T.-j. Chen, M.~Fasiello, E.~A. Lim, and A.~J. Tolley, {\it {Higher derivative
  theories with constraints: Exorcising Ostrogradski's Ghost}},  {\em JCAP}
  {\bf 1302} (2013) 042, [\href{http://xxx.lanl.gov/abs/1209.0583}{{\tt
  arXiv:1209.0583}}].

\bibitem{Malik2009}
K.~A. Malik and D.~Wands, {\it {Cosmological perturbations}},  {\em Phys.Rept.}
  {\bf 475} (2009) 1--51, [\href{http://xxx.lanl.gov/abs/0809.4944}{{\tt
  arXiv:0809.4944}}].

\bibitem{Peeters:2007wn}
K.~Peeters, {\it {Introducing Cadabra: A Symbolic computer algebra system for
  field theory problems}},  \href{http://xxx.lanl.gov/abs/hep-th/0701238}{{\tt
  hep-th/0701238}}.

\bibitem{DBLP:journals/corr/abs-cs-0608005}
K.~Peeters, {\it A field-theory motivated approach to symbolic computer
  algebra},  {\em CoRR} {\bf abs/cs/0608005} (2006).

\end{thebibliography}

\providecommand{\href}[2]{#2}\begingroup\raggedright\endgroup

\end{document}